\documentclass[10pt]{IEEEtran}

\usepackage{amsmath,amssymb}

\usepackage{subfigure}
\usepackage{graphicx}
\usepackage{float}
\usepackage{caption}
\usepackage{slashbox}
\usepackage{longtable}
\usepackage{rotating}
\usepackage{multirow}
\usepackage{array}
\usepackage{bm}
\usepackage{algorithmic}

\usepackage[ruled,vlined,linesnumbered]{algorithm2e}
\usepackage[usenames]{color}
\usepackage{amscd}
\usepackage{multirow}
\usepackage{bigdelim}
\usepackage{caption}
\usepackage{ragged2e}
\usepackage{indentfirst}
\usepackage{setspace}
\usepackage{epstopdf}
\begin{document}
%
\title{Direct Acyclic Graph-based Ledger for Internet of Things: Performance and Security Analysis}

\author{Yixin Li, Bin Cao$^*$, Mugen Peng, Long Zhang, Lei Zhang, Daquan Feng and Jihong Yu
\thanks{Y. Li (\texttt{liyixinggg@163.com}), B. Cao$^*$ (the corresponding author: \texttt{caobin65@163.com}) and M. Peng (\texttt{pmg@bupt.edu.cn}) are with the Beijing University of Posts and Telecommunications, State Key Laboratory of Networking and Switching Technology, Beijing 100876, China. Y. Li is also with the Chongqing University of Posts and Telecommunications, Chongqing 400065, China. L. Zhang (\texttt{zhanglong3211@yeah.net}) is with the University of Electronic Science and Technology of China, Chengdu 611731, China. L. Zhang (\texttt{lei.zhang@glasgow.ac.uk}) is with the School of Engineering, University of Glasgow, Glasgow, G12 8QQ, U.K. D. Feng (\texttt{fdquan@szu.edu.cn}) is with the Shenzhen University, Shenzhen 518060, China. J. Yu (\texttt{jihong-yu@hotmail.com}) is with the Beijing Institute of Technology, Beijing 100081, China.}
\thanks{This work was supported in part by the State Major Science and Technology Special Project (2018ZX033001023), in part by the National Program for Special Support of Eminent Professionals, in part by the National Natural Science Foundation of China under Grant 61701059 and Grant 61831002, in part by the Fundamental Research Funds for the Central Universities of New Teachers Project, in part by the Chongqing Technological Innovation and Application Development Projects (cstc2019jscx-msxm1322), in part by the U.K. EPSRC (EP/S02476X/1), and in part by the Zhejiang Lab’s International Talent Fund for Young Professionals.}}


\IEEEtitleabstractindextext{%
\begin{abstract}
\justifying
Direct Acyclic Graph (DAG)-based ledger and the corresponding consensus algorithm has been identified as a promising technology for Internet of Things (IoT). Compared with Proof-of-Work (PoW) and Proof-of-Stake (PoS) that have been widely used in blockchain, the consensus mechanism designed on DAG structure (simply called as DAG consensus) can overcome some shortcomings such as high resource consumption, high transaction fee, low transaction throughput and long confirmation delay. However, the theoretic analysis on the DAG consensus is an untapped venue to be explored. To this end, based on one of the most typical DAG consensuses, Tangle, we investigate the impact of network load on the performance and security of the DAG-based ledger. Considering unsteady network load, we first propose a Markov chain model to capture the behavior of DAG consensus process under dynamic load conditions. The key performance metrics, i.e., cumulative weight and confirmation delay are analysed based on the proposed model. Then, we leverage a stochastic model to analyse the probability of a successful double-spending attack in different network load regimes. The results can provide an insightful understanding of DAG consensus process, e.g., how the network load affects the confirmation delay and the probability of a successful attack. Meanwhile, we also demonstrate the trade-off between security level and confirmation delay, which can act as a guidance for practical deployment of DAG-based ledgers.
\end{abstract}

}

\maketitle

\IEEEdisplaynontitleabstractindextext

%
\IEEEpeerreviewmaketitle


%
%
%
%


\section{Introduction}

\IEEEPARstart{B}{lockchain} is a distributed ledger technology for establishing trust and consensus in peer-to-peer (P2P) networks. It is originally proposed in 2009 as the fundamental technology of crypto-currency, Bitcoin \cite{1-bitcoin}. The decentralization provided by blockchain can be largely attributed to its consensus algorithm, which enables peer-to-peer trading in a distributed manner and leverages the computational power of the whole network to ensure the immutability of the stored data. As such a safe decentralization solution, blockchain has been identified as a most promising technology to support the future digital society, and attracted much attention from both industry and academia.

Recently, blockchain has shown a great potential to be used in the Internet of Things (IoT) ecosystems, such as smart vehicles \cite{6-car}, energy trading \cite{enabling}, supply chain \cite{supplychain}, and ehealth \cite{6-IoToffloading}. Blockchain comes with characteristics of decentralization, high security, interoperation, and trust-building, which can solve the problem of high cost of infrastructure and maintenance in the traditional centralized IoT systems. According to IBM report \cite{6-IBM}, to be safe, scalable and efficient, the centralized IoT cloud systems will be transformed to blockchain-enabled decentralized systems by 2025.

It is well-known that consensus algorithm plays a key role to establish a blockchain-enabled IoT system, which motivates the nodes in the network to efficiently and securely insert the new block into the chain \cite{6-consensus}. Considering the IoT systems are typically resource-limited and large-scale, the consensus algorithm adopted in IoT systems must be resource efficiency, low cost and can support high transaction throughput. To this end, we first review the main ideas of two widely used consensus algorithms in blockchain and discuss their viability for IoT systems.

\begin{bfseries} Proof-of-Work (PoW)\end{bfseries} \cite{3-Mastering}: The core idea of PoW is the competition of computational power. The miners constantly perform hash operations to compete for the right to generate the new block with a bonus. The winner is the first miner who obtains a hash value that is lower than the announced target. On the one hand, the computational complexity in PoW must be high enough for preventing forking. But on the other hand, the high computational complexity will cause high energy consumption to generate a new block.

\begin{bfseries} Proof-of-Stake (PoS)\end{bfseries} \cite{3-PoS}: Unlike PoW that relies on computing capability, coin age is used in PoS to avoid the high computational cost of hash operations. The coin age of an unspent asset is defined as its value multiplied by the time period after it was created. In PoS, a higher coin age will result in a higher probability to obtain the right of creating a new block, and in turn the coin age would be consumed (reset as zero) when the winner receives a bonus. Since the probability to win is directly determined by coin age, PoS is beneficial to the wealthy miner, and might cause monopoly, which may result in the generation of powerful third party.

Both PoW and PoS work on a ``single chain" structure, where forking is illegal \cite{3-Mastering}. To reduce the probability of forking and maintain a single version of blockchain ledger among all nodes in the network, the consensus algorithm must slow down the generation rate of new blocks. This design principle causes the following two bottlenecks: (i) \emph{Throughput limitation}: since the capacity of the blocks is limited, the transaction throughput is usually limited to dozens, e.g., $7$ TPS in Bitcoin \cite{1-bitcoin} and $20$ to $30$ TPS in Ethereum \cite{6-ethereum}, which is unable to respond to the exponential growth of IoT nodes and needs. (ii) \emph{Confirmation delay}: low block generation rate
results in long confirmation delay, e.g., $60$ minutes in Bitcoin and $3$ minutes in Ethereum. Besides this, to maintain the security of the single chain structure, one block needs to contain high computational power or coinage. This causes the other two bottlenecks for IoT systems: (iii) \emph{Fairness}: only the nodes with high computational power or coinage have the right to generate new block. This feature cannot meet the needs of IoT systems, where the computational power of IoT nodes is usually very limited, and it is costly to provide enough token for each node. (iv) \emph{High Transaction fee}: unfairness leads to professional and powerful miners. It is a heavy burden to feedback the miners in the IoT systems with frequent micropayments.

 \begin{figure}[t]
 \setlength{\abovecaptionskip}{0.cm}
\setlength{\belowcaptionskip}{-0.4cm}
\captionsetup{font={footnotesize}}
\begin{center}
 \includegraphics[width=5.5cm]{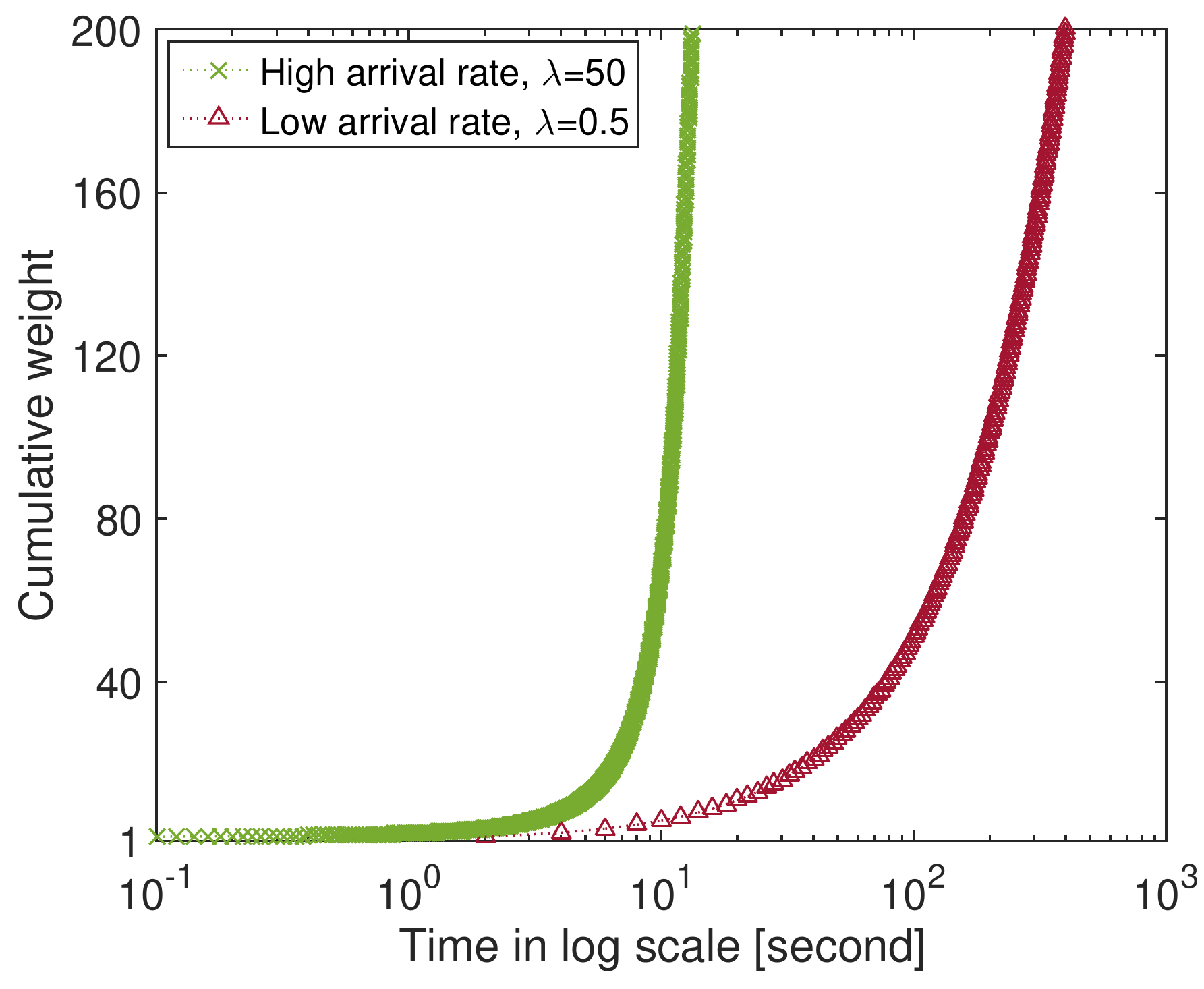}
 \end{center}
 \caption{Consensus process of a new transaction}
\label{Fig1}
\end{figure}

To overcome the above shortcomings of PoW and PoS, DAG consensus is originally proposed in \cite{7-DAG} and allows any node to insert a new block into the ledger immediately, as long as they process the earlier transactions. In this way, many forkings would be generated simultaneously. This phenomenon is regarded as a problem in many traditional consensus process since it would cause ``double-spending" \cite{8-double-spend}. However, DAG consensuses use some effective algorithms (e.g., Markov Chain Monte Carlo algorithm \cite{9-tangle} and virtual voting algorithm \cite{11-hashgraph}) to address double-spending problem and allow new arrival transactions access the ledger in a forking topology. As a result, the transaction throughput in DAG consensus process will not be limited anymore \cite{15-comparsion}. Moreover, unlike the single chain design in PoW or PoS, the data stored in DAG-based ledgers is protected by massive forking blocks, thus, the average resource consumption on each node can be very low. Accordingly, the professional miner is not necessary and low or no transaction fee is possible, which is critical to IoT ecosystems.

Despite many advantages for IoT, DAG consensus also faces some challenges. In practical IoT systems, it is impossible that new transactions arrive quickly and steadily all the time. When the transaction arrival rate becomes low, the confirmation delay will increase significantly since the earlier transactions must wait for the new transactions to process. In \cite{9-tangle}, the growth curves of cumulative weight in high load and low load regimes are analysed, which are shown in Fig. 1, where $\lambda$ represents the transaction arrival rate (transactions per second). The cumulative weight of a transaction is proportional to the amount of computational power accumulating behind it. When the cumulative weight reaches the defined threshold, the transaction is confirmed and the probability of malicious modification is very low. We can see that the growth rate of cumulative weight in the low load regime is much lower than that in the high load regime, which will result in a long confirmation delay. In fact, the network load is determined by transaction arrival rate which could be fluctuant in practical IoT systems. In such an unsteady load regime, the performance of DAG consensus process becomes more complicated. Moreover, the network load will also affect the security of system, where a lower load will result in a lower cumulative weight growth rate, and thus lead to a higher probability of a successful double-spending attack.


Inspired by these observations, this paper aims to investigate the impact of network load on the performance and security of DAG consensus in an analytical manner. First, we introduce a Markov chain model to capture the impact of network load on the performance of DAG consensus process in terms of cumulative weight growth rate and confirmation delay. Then, we formulate attack strategies and leverage a stochastic model to examine the probability of a successful double-spending attack in different network load regimes. The analytical models and results can provide an insightful understanding of the performance and security in the DAG consensus. The main contributions of this paper can be summarized as follows.
\begin{itemize}
\item We point out the impact of network load on the performance and security of DAG consensus. By classifying four network load regimes, we reflect this impact in a qualitative and quantitative manner.

\item Considering the characteristics of fluctuant network load in practical IoT systems, we propose Markov chain model for DAG consensus process and capture the impact of the change in load mathematically. The proposed model demonstrates the relationship between the action of nodes in DAG network and the corresponding influence to system performance, which offers an insightful observation of DAG consensus process.

\item We examine the attack strategy based on network load using a stochastic model, and derive the expression of the probability to conduct a successful double-spending attack. The equations can indicate the required computational power of attacker for double-spending in different load regimes. This analysis clearly explains the malicious action of attacker, and thus serves as a theoretical guidance to protect the honest transactions.
\item Through extensive experiments, we validate our analysis and obtain insightful results: (i) compared with the steady high load regime, when the network load changes from high to low, the confirmation delay will increase significantly (in worst case, it is even much longer than low load regime). In contrast, when the load changes from low to high, the confirmation can happen very fast. (ii) the adaptation period (introduced in section IV) in consensus process can be used to increase the probability of a successful attack. (iii) the trade-off between security level and confirmation delay can provide a guideline to find a suitable confirmation threshold for protocol design of DAG consensus.
\end{itemize}

The rest of this paper is organized as follows. Section II provides some basic principles in the DAG-based ledger. In Section III, we introduce the Markov chain model for consensus process. Based on the proposed model, Section IV analyses the performance in terms of cumulative weight growth and confirmation delay under different network load regimes. Section V introduces the double-spending attack in the DAG-based ledger, and use a stochastic model to study the attack process. In Section VI, we examine the attack strategy in DAG consensus process and obtain the probability of a successful attack under different network load regimes. Section VII conducts some experiments for comparisons and discussions. Section VIII reviews some related work, and finally, Section IX concludes the whole paper.

\section{Preliminaries}

\subsection{The Basic Principles}

The principle of DAG consensus is to attach the new transactions in a forking topology. Under such design, there are several proposed consensus algorithms, such as Tangle \cite{9-tangle}, Byteball \cite{10-byteball} and Hashgraph \cite{11-hashgraph}. Among them, Tangle is the first proposed one that has attracted much attention in IoT field, and it has the highest market capitalization in DAG-based ledgers \cite{12-MarketCap}. Therefore, we adopt it as a typical example to examine DAG consensus process in this work.


\begin{figure}[t]
 \setlength{\abovecaptionskip}{0.cm}
\setlength{\belowcaptionskip}{-0.4cm}
\captionsetup{font={footnotesize}}
\begin{center}
\includegraphics[width=7cm]{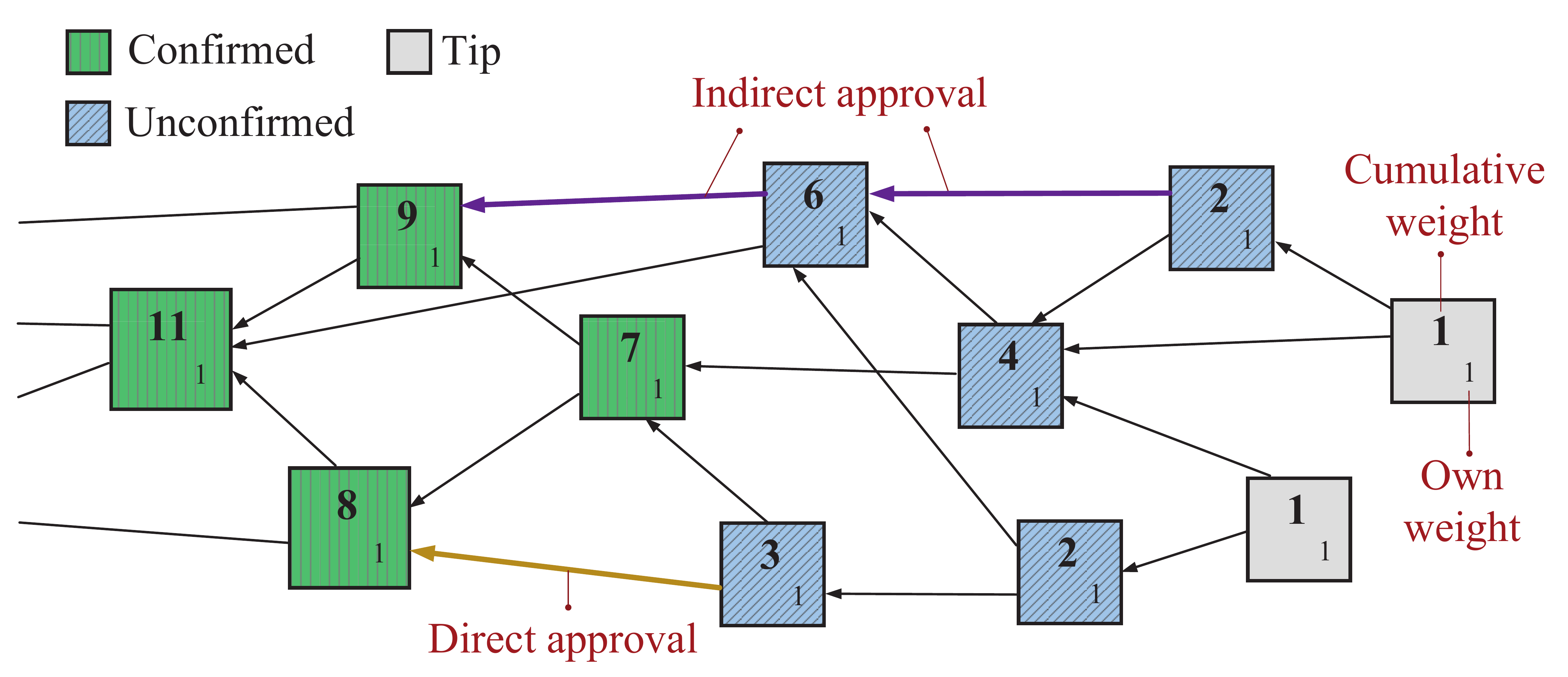}
\end{center}
\caption{An example of consensus process in Tangle
}
\label{Fig2}
\end{figure}

Tangle is the mathematical foundation of IOTA, a cryptocurrency for the IoT industry \cite{9-tangle}. As a DAG-based ledger for recording transactions, Tangle allows different branches to eventually merge into the chain, and thus leads to a much higher overall throughput compared with PoW and PoS. To access the DAG-based ledger as a new block, any new transaction has to approve a number of earlier transactions (typically two \cite{9-tangle}). Thanks to this, the higher transaction arrival rate, the faster a transaction can be confirmed. Moreover, since the workload to create a new block is light, the powerful professional miners are not necessary in this network. As a result, all nodes can issue their own transactions without a transaction fee. This is critical to the IoT applications, since micro-payments are typical trading scenarios. Some basic concepts in Tangle are listed as follows, and also illustrated in Fig. 2.

\begin{bfseries}\emph{Block}\end{bfseries}: all the blocks in Fig. 2 are the storage units to record information including transaction, digital signature, and hash value. Since one block records one transaction in Tangle, a block can be simply called as a transaction. \begin{bfseries}\emph{Tip}\end{bfseries}: it is the transaction (or block) that has not been approved yet. \begin{bfseries}\emph{Direct approval}\end{bfseries} and \begin{bfseries}\emph{indirect approval}\end{bfseries}: as shown in Fig. 2, each edge represents an approval, a direct edge indicates the direct approval, and a path between two transactions with multi-hop indicates the indirect approval. \begin{bfseries}\emph{Own weight}\end{bfseries}: the own weight of a transaction is proportional to the amount of work which is put in by its issuer. \begin{bfseries}\emph{Cumulative weight}\end{bfseries}: it is the sum of a transaction's own weight and the overall own weights of the transactions that directly or indirectly approve it. Cumulative weight stands for the confirmation level of a transaction in the DAG-based ledger.


\begin{figure}[t]
\setlength{\abovecaptionskip}{0.cm}
\setlength{\belowcaptionskip}{-0.4cm}
\captionsetup{font={footnotesize}}
\begin{center}
\includegraphics[width=8.5cm]{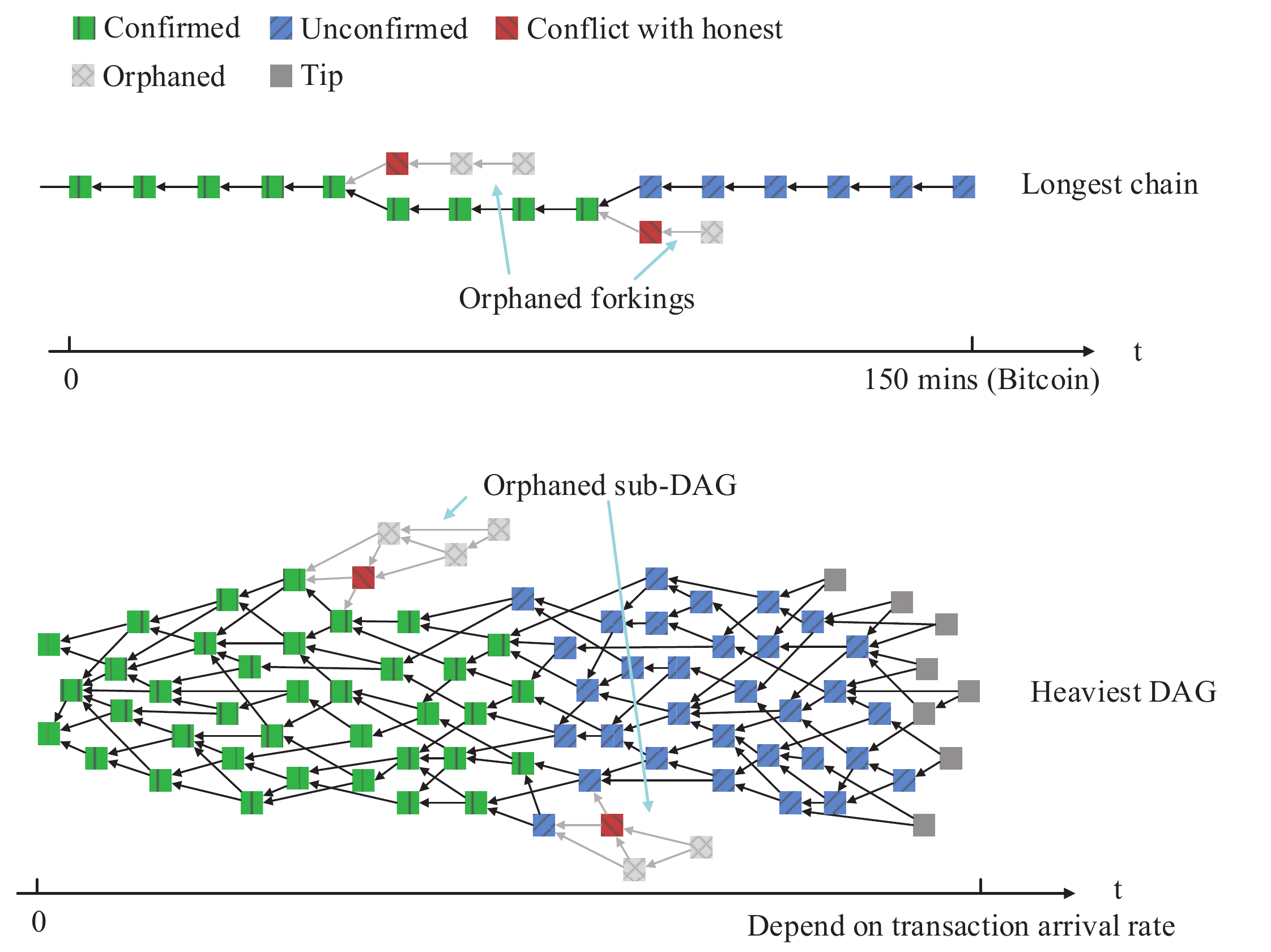}
\end{center}
\caption{Longest chain in PoW vs. Heaviest DAG in Tangle}
\label{Fig2}
\end{figure}

\subsection{Consensus Process}

To issue a new transaction and let the other nodes accept it (i.e., reach an agreement for the consensus), the main procedures are listed as follows. (i) A node creates a storage unit to store the new transaction. (ii) The node selects two tips with no-conflict according to Markov Chain Monte Carlo (MCMC) tips selection algorithm \cite{9-tangle}, and adds the hash of the selected tips into its storage unit. (iii) The node finds a nonce to solve a cryptographic puzzle to meet the difficulty target, which is similar to PoW but with a very low difficulty-of-work for avoiding spamming. (iv) The node uses its private key to sign the new transaction and broadcasts it to others. (v) When the other nodes receive it, they check whether it is legal or not based on the digital signature and nonce. For simplicity of later analysis, we define procedures (i) to (v) as the reveal stage of a new transaction.

After that, the successfully checked new transaction will be added as a new tip in the DAG-based ledger, and then wait for confirmation through direct approval and indirect approval of subsequent transactions till its cumulative weight reaches the defined threshold. This process is defined as the weight accumulation stage of a new transaction.

\begin{figure*}
\begin{equation}\label{tips}\small
\underbrace{\frac{\lambda h_{r}(\lambda h_{r}\!-\!1)}{(r+\lambda h_{r})(r+\lambda h_{r}-1)}\!\times\!0\!}_{\textrm{L(t)+1}}+\underbrace{\!\frac{2r\lambda h_{r}}{(r+\lambda h_{r})(r+\lambda h_{r}-1)}\!\times\!1\!}_{\textrm{L(t)+0}}+\underbrace{\!\frac{r(r-1)}{(r+\lambda h_{r})(r+\lambda h_{r}-1)}\!\times\!2\!}_{\textrm{L(t)-1}}=\!\frac{2r}{r+\lambda h_{r}}
\end{equation}
\vspace*{-16pt}
\end{figure*}

\subsection{Forking Problem and the Solution}

In a distributed ledger, building forking to redo the work is the only way to tamper with the stored data. Based on this, the main idea of the double-spending attack is to place two conflicting transactions at two chains in parallel. After the first transaction is spent on service, the attacker extends the chain containing the conflicting transaction and lets it outpace the first chain. When this action succeeds, the first transaction will be orphaned and the attacker can spend the token more than once.

To address forking for security, the single chain-based ledger (e.g., Bitcoin) uses the longest chain as the criterion, which is shown in Fig. 3. To maximize its profit, a rational miner should work on the longest chain when forking occurs, since the longest chain has the lowest probability to be orphaned \cite{3-Mastering}. In the DAG-based ledger, although DAG topology can support a high performance in consensus process, the forking in DAG also should be limited to a reasonable scale for preventing double-spending. Similar to Bitcoin, IOTA uses the heaviest DAG to address forking problem (sub-DAG). To this end, a rational node in DAG network should use the MCMC tip selection algorithm to extend the heaviest DAG, which has the highest overall cumulative weight. Meanwhile, the sub-DAG with less overall cumulative weight will not be approved by new transactions gradually. In summary, both the honest miners in Bitcoin and the honest nodes in IOTA use their own computational power to prevent data from tampering.

\section{Markov Chain Model for Consensus Process}

In this section, we propose a Markov chain model to analyse the consensus process of an observed transaction under unsteady network load regimes.

\subsection{System Model}

Recalling that we have divided the consensus process of an observed transaction in a DAG-based ledger into two stages: reveal stage and weight accumulation stage. Reveal stage is to attach the observed transaction into the DAG-based ledger, so that the transaction can be seen by all nodes. Let the average duration time in reveal stage be $h_r$, which is determined by the computation and transmission time. In weight accumulation stage, the cumulative weight of the observed transaction increases from its own weight to confirmation threshold (denoted by $m$) gradually. Without loss of generality, we normalize the average own weight of each transaction into $1$, and thus the cumulative weight of the observed transaction is $1$ plus the overall number of transactions that directly or indirectly approve it.

Considering the nodes of a DAG-based ledger are roughly independently distributed in a large scale IoT network, it is reasonable to assume the new transaction arrival follows Poisson process. Let $\lambda$ be the arrival rate of the new transactions issued by the honest nodes. When a new transaction arrives, it will select two tips using MCMC algorithm. The principle of MCMC algorithm is to independently place some particles on the old transactions of the DAG-based ledger and let these particles perform random walks towards the tips. To orphan the sub-DAG, these particles prefer to go through the transactions with a higher cumulative weight. Since the difference of cumulative weight among neighbouring transactions in the heaviest DAG is very small, we can approximatively consider that each tip in the heaviest DAG can be randomly selected by MCMC algorithm with equal probability. On the other hand, the overall cumulative weight of the heaviest DAG is much larger than that of the sub-DAG, so that MCMC algorithm will choose tips in the heaviest DAG and the sub-DAG generated by attacker will be orphaned.

Moreover, to analyse the impact of network load, we classify the network load into four regimes: High load Regime (HR), Low load Regime (LR), High to Low load Regime (H2LR) and Low to High load Regime (L2HR) as follows.

\subsection{Steady Regime: HR}

The network load (transaction arrival rate) keeps steady in this regime. Let $h\!=\!1/\lambda$ be the average interarrival time between two transactions. When $h\!\leq\!h_r$, it means that the network load is high, and it is defined as HR. In the DAG-based ledger, after a new transaction directly approves two tips, it will be a new tip and the selected two will be covered (they are no longer tips and the other incoming transactions should not directly approve them). However, when $h\!\leq\!h_r$, many new transactions would arrive at the reveal stage of earlier transactions, and the tips selected by earlier transactions have not been broadcast to the network. As a result, it is probable that the same tip will be directly approved by several different transactions, and thus the number of tips will keep steady, intuitively.

Let $L(t)$ be the number of tips in the heaviest DAG at time $t$. According to the analysis in \cite{9-tangle}, $L(t)$ fluctuates around a constant value $L$.
Based on the stability of tips, we have $L(t)\!=\!L(t\!-\!h_{r})\!=\!L$. Meanwhile, we know that there are $\lambda h_{r}$ new transactions arrive during $t\!-\!h_{r}$ to $t$ on average. As a result, at time $t$, $\lambda h_{r}$ new tips in $L(t)$ will replace $\lambda h_{r}$ old tips in $L(t\!-\!h_{r})$. Therefore, we can rewrite $L(t)\!=\! r \!+\!\lambda h_{r}$, where $r$ represents the old tips, and $\lambda h_{r}$ represents the tips chosen by the new transactions during $t\!-\!h_{r}$ to $t$ (they are not tips anymore, but other nodes do not know at this time).

Moreover, when a new transaction arrives at time $t$, it would select tips randomly from $L(t)$. Since $\lambda h_r$ are not tips anymore, tips selection from $\lambda h_r$ or $r$ will affect the number of $L(t)$ in the future. If the new transaction selects zero tip in $r$, $L(t)$ will increase by $1$; if it selects one tip in $r$, $L(t)$ will remain unchanged; otherwise, $L(t)$ will decrease by $1$. The expected number of selected tips in $r$ can be calculated in (\ref{tips}). Based on the stability of $L(t)$, we have $\frac{2r}{r+\lambda h_{r}}\!=\!1$. Thus, $r\!=\lambda h_{r}$ and $L\!=\!L(t)\!=\!2\lambda h_r$.




\subsection{Steady Regime: LR}

Compared with HR, LR is the situation when $h\!>\!h_{r}$. In this case, when a new transaction arrives, the earlier transactions have revealed to the DAG-based leger in expectation. Since one transaction covers two tips, the typical number of tips in this regime will decline, and becomes $1$ finally. Note that $L\!=\!2\lambda h_r$ is also available in LR, where $L\!=\!2\lambda h_r\!\approx\!1$ based on $h\!>\!h_{r}$.

\subsection{Unsteady Regime: H2LR}

The consensus process of an observed transaction in HR and LR have been explored in \cite{9-tangle}. In this work, we focus on the consensus process in unsteady regimes. The transaction arrival rate is steady in HR and LR, which can be denoted by $\lambda_{h}$ and $\lambda_{l}$, respectively. When the transaction arrival rate changes from $\lambda_{h}$ to $\lambda_{l}$ suddenly, it is an unsteady regime and defined as H2LR. Accordingly, the number of tips will decrease from $2\lambda_{h} h_r$ (denoted by $L_{h}$) to $2\lambda_{l} h_r\!=\!1$ gradually.

As a metric of confirmation level, let $W(t)$ be a stochastic process representing the cumulative weight of an observed transaction at time $t$. It will increase with the approval of new transactions over time. Meanwhile, the probability to approve the observed transaction is affected by the number of tips $L(t)$ based on random selection, and $L(t)$ is also a stochastic process. Therefore, when the transaction arrival rate becomes low, we can have the value of $\{W(t),L(t)\}$ at the next moment only depends on the present and is independent of the past. Furthermore, when the transaction arrival rate is low, we can approximatively consider that the transactions attach to the DAG-based ledger one by one. 
Therefore, $\{W(t),L(t)\}$ can be formulated as a discrete-time Markov chain $\{W(k),L(k)\}, k\!=\! 0, 1, 2, . . .\infty $, where the state changes with the arrival of each new transaction.

\begin{figure}[t]
\setlength{\abovecaptionskip}{0.cm}
\setlength{\belowcaptionskip}{-0.4cm}
\captionsetup{font={footnotesize}}
\begin{center}
 \includegraphics[width=8.3cm]{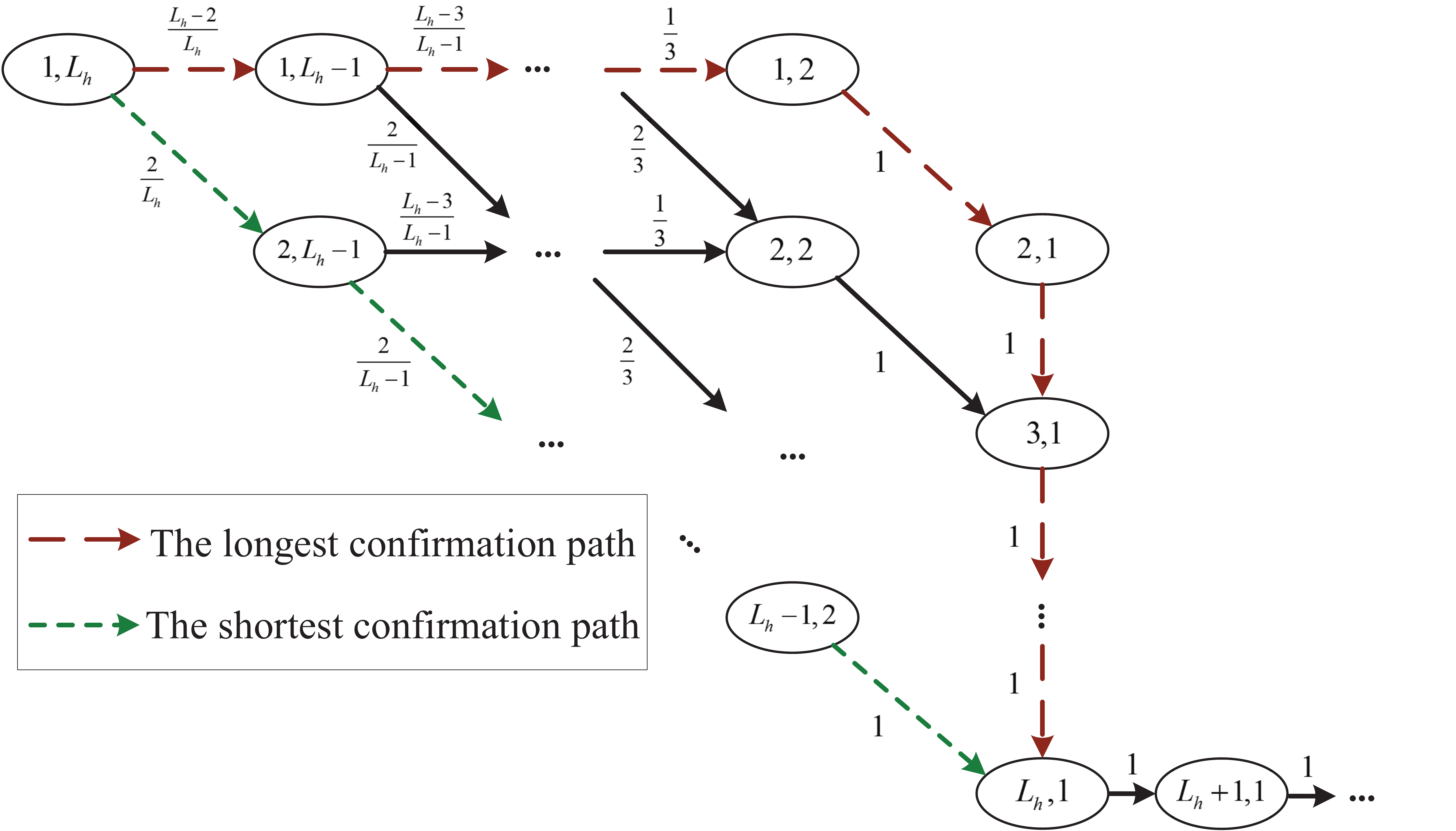}
 \end{center}
 \caption{{Markov Chain model for the consensus process of an observed transaction under H2LR.}}
\label{Fig2}
\end{figure}

The Markov chain model for an observed transaction under H2LR is shown in Fig. 4. The initial state represents that the observed transaction reveals to the DAG-based ledger under HR, where $W(0)\!=\!1$, $L(0)\!=L_{h}\!=\!2\lambda_{h} h_r$. The observed transaction is confirmed when $W(k)\!=\!m$, where $m\!\geq\!2$. In each new transaction arrival interval, $W(k)$ of the observed transaction will remain the same or increase by $1$ based on the result of random selection. Since the new transaction will choose two tips from $L(k)$ randomly, the probability to select the observed transaction for $W(k\!+\!1)\!=\!W(k)\!+\!1$ is $\frac{1}{L(k)}\!\times\!\frac{L(k)-1}{L(k)-1}\!+
\!\frac{L(k)-1}{L(k)}\!\times\frac{1}{L(k)-1}\!=\!\frac{2}{L(k)}$. Accordingly, the probability of not being selected for $W(k\!+\!1)\!=\!W(k)$ is $\frac{L(k)-2}{L(k)}$. When the new transaction approves two tips, it will be a new tip and the selected two are not tips anymore. In this case, $L(k)$ will decrease by $1$ in each arrival interval until $L(k)\!=\!1$. Especially, when $L(k)$ reduces to $2$, the observed transaction will be approved by the incoming transaction with probability $1$, and thus $W(k)$ will increase by $1$ and $L(k)$ will decrease by $1$. In the following, $L(k)$ remains $1$ and $W(k)$ increases linearly with speed $\lambda_{l}$. Based on above analysis, the one-step transition probabilities can be given by
\begin{equation}\label{4}\small
\begin{split}
\begin{cases}
P\left\{ {i\!+\!1,j\!-\!1\mid i,j} \right\} \!=\! 2/j,~~i\!=\!1,2,\cdots, L_{h}\!-\!1;\,j \!=\!2,3,\cdots, L_{h},\\
P\left\{ {i,j \!-\! 1\mid i,j} \right\} \!=\! 1\!-\!2/j,~~i\!=\!1,2,\cdots, L_{h}\!-\!1;\, j\!=\!2,3,\cdots, L_{h},\\
P\left\{ {i \!+\! 1,1\mid i,j} \right\} \!=\! 1,~~~~~ \, ~~~i\!=\!2,3,\cdots,\infty;~~~~ \, \, j\!=\!1.
\end{cases}
\end{split}
\end{equation}
We adopt the short notation, where {\small$P\left\{ {i\!+\!1,j\!-\!1\mid i,j} \right\} \!=\!P\left\{ {W(k\!+\!1)\!=\!i\!+\!1,L(k\!+\!1)\!=\!j\!-\!1\mid W(k)\!=\!i,L(k)\!=\!j} \right\}$.}


The first equation in (\ref{4}) stands for the situation that the observed transaction has been approved by an incoming new transaction, thus $W(k\!+\!1)\!=\!W(k)\!+\!1$ and $L(k\!+\!1)\!=\!L(k)\!-\!1$. The second equation stands for the situation that the observed transaction has not been approved, so $W(k\!+\!1)\!=\!W(k)$ and $L(k\!+\!1)\!=\!L(k)\!-\!1$. The third indicates that H2LR has transferred to LR. The observed transaction will be approved by the following new transactions with probability $1$, since it has been indirectly approved by all tips.

Note that the above discussion is based on the worst case to study the lower performance bound in H2LR, where transaction arrival rate changes from $\lambda_{h}$ to $\lambda_{l}$ as soon as the observed transaction reveals in the network and $W(0)\!=\!1$. In contrast, the best case for upper performance bound in this regime is that transaction arrival rate changes from $\lambda_{h}$ to $\lambda_{l}$ when $W(k)\!=\!m$, which is similar to the consensus process under HR. As an extending, the transaction arrival rate can change from high to low at the any state of the observed transaction by integrating the analysis of HR in \cite{9-tangle} and the proposed Markov chain model in H2LR.

\subsection{Unsteady Regime: L2HR}

\begin{figure}[t]
\setlength{\abovecaptionskip}{0.cm}
\setlength{\belowcaptionskip}{-0.4cm}
\captionsetup{font={footnotesize}}
\begin{center}
 \includegraphics[width=4.5cm]{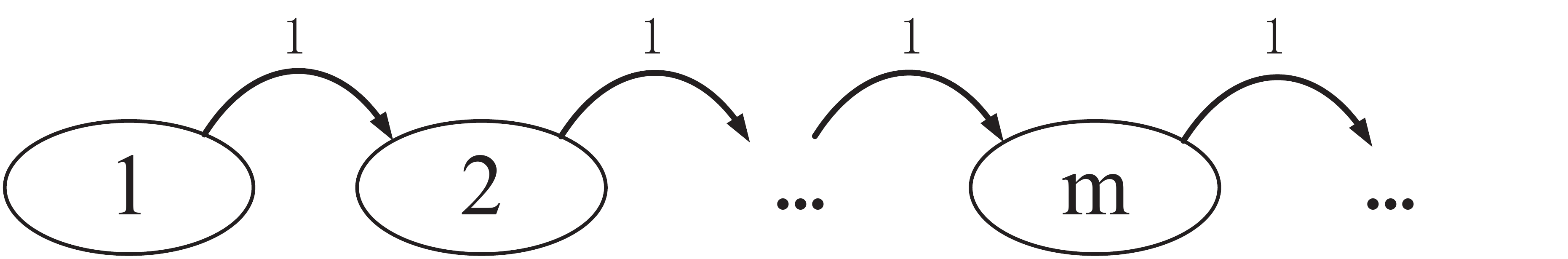}
 \end{center}
 \caption{{Markov Chain model for the consensus process of an observed transaction under L2HR.}}
\label{Fig3}
\end{figure}

Compared with H2LR, L2HR happens when the arrival rate increases from $\lambda_l$ to $\lambda_h$. Accordingly, the number of tips increases from 1 to $2\lambda_h h_r$ gradually.

The Markov chain model for an observed transaction under L2HR is illustrated in Fig. 5. Since the observed transaction reveals under LR where $L(0)\!=\!1$, it is fully covered and will be directly or indirectly approved by all the new transactions. As a result, $W(k)$ will increase linearly with speed $\lambda_h$ regardless of $L(k)$. The transition probabilities under L2HR are shown as follows.
\begin{equation}\label{6}\small
P\left\{ {W(k\!+\!1)\!=\!i \!+\!1\mid W(k)\!=\!i} \right\} \!=\! 1,
\end{equation}
where $i\!=\! 1, 2, . . .\infty$.

Similarly, we use this model to capture the best case to study the upper performance bound in L2HR where transaction arrival rate changes from $\lambda_{l}$ to $\lambda_{h}$ when $W(0)\!=\!1$. In contrast, the worst case for lower performance bound in L2HR is that transaction arrival rate changes from $\lambda_{l}$ to $\lambda_{h}$ when $W(k)\!=\!m$, which can refer to the consensus process under LR.

Note that the unsteady regimes defined in this work refer to abrupt changes of load to provide upper and lower performance bounds. For the slow-changing load case, we can approximately predict its impact on performance by using the derived performance bounds. However, the closed-form expression for the slow-changing load case is not analytically derived in this work.

\section{Performance Analysis}


\subsection{Cumulative Weight}

\textbf{HR:} The growth of cumulative weight under steady regimes, HR and LR, has been discussed in previous work \cite{9-tangle}. We briefly review this work as the preliminaries to provide further analysis of confirmation delay and double-spending. The cumulative weight of an observed transaction begins to grow when the reveal stage ends. In HR, the weight accumulation stage has two periods: adaptation period and linear growth period. The adaptation period of an observed transaction can be thought as the time until almost all the tips indirectly approve that transaction. The expected cumulative weight of an observed transaction grows with $E[W(t)] \!=\! 2\exp(0.352t/{h_{r}})$ during adaptation period \cite{9-tangle}. Next, when the adaptation period ends, all incoming transactions will indirectly approve the observed transaction, and the expected cumulative weight grows with speed $\lambda_{h}$, which is called as linear growth period. Let $t_{0}$ be the duration time of adaptation period. The adaptation period ends when cumulative growth rate becomes $\lambda_{h}$, namely $\frac{{dE[W(t)]}}{{dt}}|{_{t = {t_0}}}\! = \!{\lambda _h}$. Accordingly, we can obtain $t_{0}\!=\!\frac{h_{r}}{0.352}\!\times \!ln(L_{h}/{1.408})$ and $W(t_{0})\!=\!\frac{L_{h}}{0.704}$. Hence, the cumulative weight growth of an observed transaction in this regime is
\begin{equation}\label{HR_w}\small
E[W(t)_{hr}]\! =\! \left\{ {\begin{array}{*{20}{l}}
{2\exp (0.352t/{h_r}),{\rm{  }}~~~~~0\!\leq \!t\!\leq \!t_{0}},\\
{\frac{L_{h}}{0.704}\! +\! {\lambda_h}(t \!-\! {t_0}),{\rm{     }}~~~~~~~~~~ t \!>\! t_{0}. }
\end{array}} \right.
\end{equation}

\textbf{LR:} Since $L(0) \!= \!1$ in LR, the incoming new transactions will approve the observed transaction with probability $1$. Consequently, the average cumulative weight growth rate is $\lambda_{l}$ in this regime. The expected cumulative weight in LR at time $t$ can be expressed as
\begin{equation}\label{LR_w}\small
E[W(t)_{lr}]\! =\! 1\!+\!{\lambda _l}t,~~t \!\geq\! 0.
\end{equation}

\textbf{H2LR:} As shown in Fig. 4, when $0\!\leq\!k\!\leq \!L_{h}\!-\!1$, each column of the state transition diagram stands for all possible states $\{W(k), L(k)\}$ at a specific step $k$. For example, when $k\!=\!0$, the possible state is $\{1,L_{h}\}$; when $k\!=\!1$, the possible states are $\{1,L_{h}\!-\!1\}$ and $\{2,L_{h}\!-\!1\}$; when $k\!=\!L_{h}\!-\!1$, the possible states are $\{2,1\}, \{3,1\} ,. . ., \{L_{h},1\}$. In the case of $k\!\ \geq\!L_{h}$, the number of possible states will remain $L_{h}\!-\!1$. For example, if the step moves from $L_{h}\!-\!1$ to $L_{h}$, the cumulative weight of all possible states will increase by $1$ simultaneously, i.e., change from $\{2,1\}, \{3,1\} ,. . ., \{L_{h},1\}$ to $\{3,1\}, \{4,1\} ,. . ., \{L_{h}\!+\!1,1\}$. The reason is that the observed transaction has been indirectly approved by all tips when $k\!\ \geq\!L_{h}$.

Based on this, we could obtain the expected cumulative weight at step $k$ in H2LR as 
\begin{equation}\label{h2lr}\small
\begin{split}
E[W(k)_{h2lr}] = & \sum\limits_{\forall i} \sum\limits_{\forall j} P\{ W(k) = i,L(k) = j \mid \\
&{W(0) = 1,L(0) = {L_h}} \}   \times i,
\end{split}
\end{equation}
where $k \!=\! 0, 1, \cdots ,\infty$, and $P\{ W(k)\! =\! i,L(k)\! = \!j \!\mid \!{W(0) \!= \!1,L(0) \!= \! {L_h}} \}$ is the $k$-step transition probability which can be calculated from (2). If and only if $\{i,j\}$ is a possible state at step $k$, the corresponding $k$-step transition probability is greater than $0$.

As mentioned before, the new transaction arrival is a Poisson process. Let $\{H_{i}, i\!=\! 1, 2, . . .\infty \}$ be the sequence of interarrival times between two neighboring transactions, where $H_{i}, i \!=\! 1, 2, . . .\infty,$ are independent and identically distributed exponential random variables with mean $1/\lambda_{l}$ under H2LR. 
According to $t\!=\!\sum\limits_{i=1}^k {{H_i}}$, (\ref{h2lr}) can be transformed as the expected cumulative weight at time $t$ as follows.
\begin{equation}\label{H2LR_w}\small
\begin{split}
E[W(t)_{h2lr}] = & \sum\limits_{\forall i} \sum\limits_{\forall j} P\{ W(t) = i,L(t) = j \mid \\
&{W(0) = 1,L(0) = {L_h}} \}   \times i,
\end{split}
\end{equation}
where $t \!=\! 0, H_{1}, H_{1}\!+\!H_{2}, H_{1}\!+\!H_{2}\!+\!H_{3}, \cdots ,\infty$.

\textbf{L2HR:} In this regime, due to $L(0) \!= \!1$, all new incoming transactions will direct and indirectly approve the observed transaction. As a result, $W(k)$ increases by $1$ with probability $1$ in each transaction arrival interval. The expected cumulative weight with $k$ in L2HR is
\begin{equation}\small
E[W(k)_{l2hr}]\! =\! 1\!+\!k,~~k \!=\! 0,1,2, \cdots ,\infty.
\end{equation}

The expected cumulative weight in L2HR at time $t$ can be expressed as
\begin{equation}\label{L2HR_w}\small
E[W(t)_{l2hr}]\! =\! 1\!+\!k,~~t\!=\!\sum\limits_{i=1}^k {{H_i}},
\end{equation}
where $t \!=\! 0, H_{1}, H_{1}\!+\!H_{2},\cdots ,\infty$.

\subsection{Confirmation Delay}

Confirmation delay is defined as the time from $W(0)\!=\!1$ to $W(t)\!=\!m$.

\begin{figure*}\small
\begin{equation}\label{H2LR_d}
E[{T_{h2lr}}] = \left\{ {\begin{array}{*{20}{l}}
{\rm {Case \ I:}} \begin{array}{l}
\sum\limits_{k = m\! -\! 1}^{{L_h} \!- \!2} {P\{ W(k\! -\! 1)\! =\! m \!-\! 1,L(k \!- \!1) \!= \!{L_h} \!-\! k \!+\! 1\left| {W(0)\! = \!1,L(0) \!=\! {L_h}} \right.\} } \! \times\! \frac{2}{{{L_h}-k+ 1}}\! \times \!k{h_l}\\
 \!+\! \sum\limits_{k = {L_h}\! -\! 1}^{m +\! {L_h} \!- \!3} {P\{ W(k) \!= \!m,L(k) \!=\! 1\left| {W(0)\! =\! 1,L(0) \!= \!{L_h}} \right.\} }\!  \times \!k{h_l},~~~~~~~~~~~~2\!\leq\!m\!<\!L_{h},
\end{array}\\
{\rm{Case \ II:}} {\sum\limits_{k = m\! - \!1}^{m +\! {L_h}\! -\! 3} {P\{ W(k)\! =\! m,L(k) \!= \!1\left| {W(0)\! =\! 1,L(0)\! =\! {L_h}} \right.\} } \! \times \!k{h_l},~~~~~~~~~~ \, ~~~~~~~~  m\!\geq\!L_{h}.}
\end{array}} \right.
\end{equation}
\vspace*{-10pt}
\end{figure*}

\textbf{HR:} Let $E[T_{hr}]$ be the expected confirmation delay in HR. Based on (\ref{HR_w}), if confirmation threshold $m\!\leq\![W(t_{0})]$, the observed transaction will be confirmed during adaptation period. Accordingly, we have $m \!=\! 2\exp (0.352E[T_{hr}]/{h_r})$. Otherwise, the confirmation will happen during linear growth period, where $m \!=\! \frac{L_{h}}{0.704}\! +\! {\lambda_h}( E[T_{hr}]\!-\! {t_0})$. We can obtain that
\begin{equation}\label{HR_d}\small
E[{T_{hr}}] = \left\{ {\begin{array}{*{20}{l}}
{\underbrace{\frac{{{h_r}}}{{0.325}}\ln(m/2)}_{\textrm{Confirmed in adaptation}},~~~~~~~~~~~~~~~~~~~~~~~~2\!\leq\!m\!\leq\![W(t_{0})],}\\
{\underbrace{\frac{h_{r}}{0.325}\ln(L_{h}/1.408)}_{\textrm{Time for adaptation}}+\underbrace{\frac{1}{{\lambda _h}}(m\!-\!\frac{L_{h}}{0.704})}_{\textrm{Time for liner growth}},m\!>\![W(t_{0})].}
\end{array}} \right.
\end{equation}

%
%

\begin{figure}[t]
\setlength{\abovecaptionskip}{0.cm}
\setlength{\belowcaptionskip}{-0.cm}
\captionsetup{font={footnotesize}}
\begin{center}
 \includegraphics[width=7.3cm]{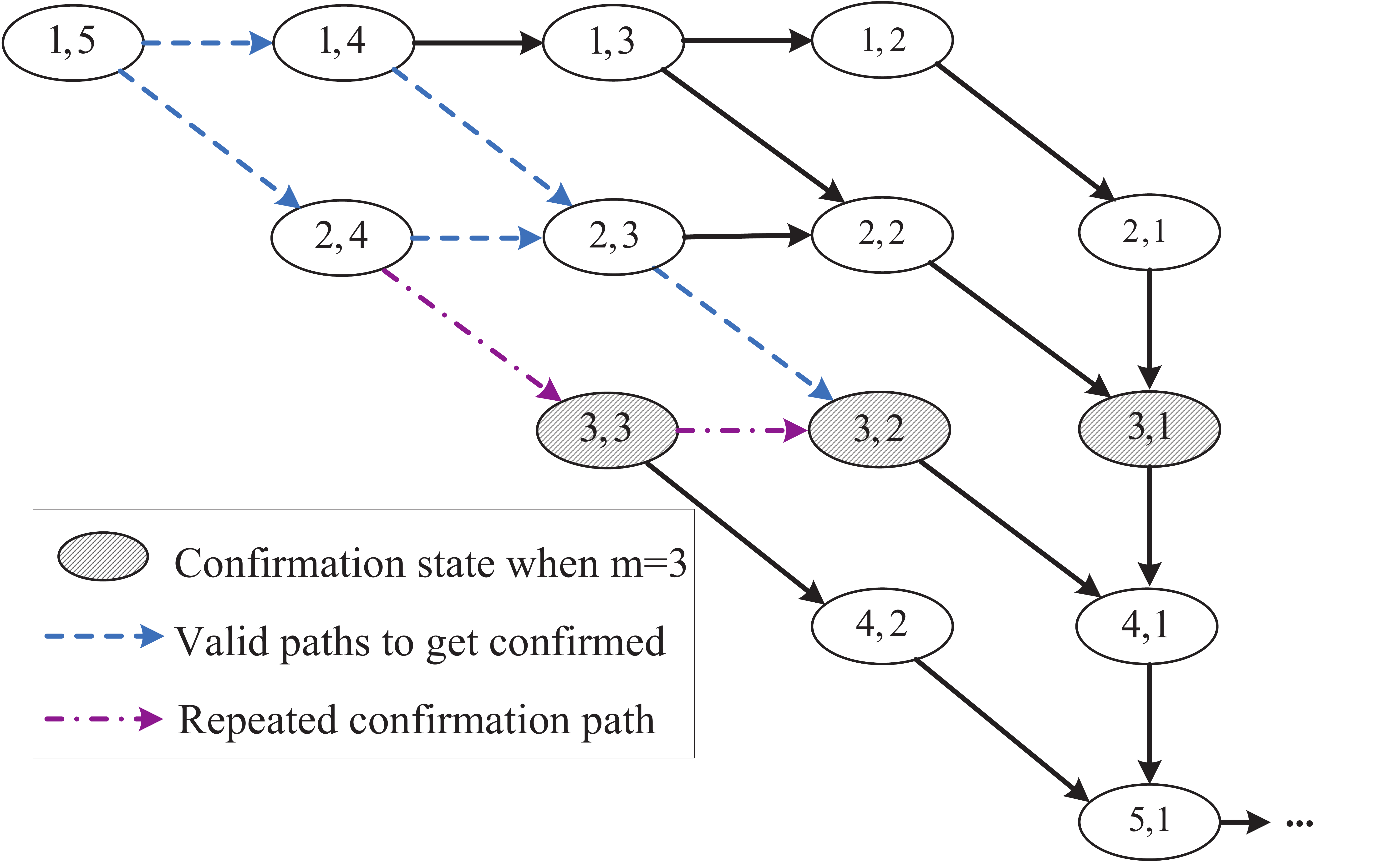}
 \end{center}
 \caption{{Simplified Markov Chain model in H2LR; $L_{h}=5$ and $m=3$}}
\label{Simplified}
\end{figure}

\textbf{LR:} Let ${E[T_{lr}}]$ be the expected confirmation delay in this regime. Based on the cumulative weight growth of LR in (\ref{LR_w}), we can obtain that
\begin{equation}\label{LR_d}
E[T_{lr}] \!=\! {(m\!-\!1)}{h_{l}},~~m\!\geq\!2,
\end{equation}
where $h_{l}\!=\!1/\lambda_{l}$.

\textbf{H2LR:} As shown in Fig. 4, there are various paths from the initial state $\{1,L_{h}\}$ to the confirmation state $\{m, L(k)\}$. Among them, the green path with short dashed is the shortest one, where the transaction will be approved by $m\!-\!1$ new incoming transactions with the smallest expected confirmation delay. In contrast, the red path with long dashed is the longest confirmation path that goes through $m\!+\!L_{h}\!-\!3$ new transactions. Let $E[T_{h2lr}]$ be the expected confirmation delay in H2LR, which can be expressed as (\ref{H2LR_d}). Note that $\frac{2}{{{L_h}-k+ 1}}$ is the probability $P\{ W(k)\! =\! m,L(k) \!=\! {L_h}\! - \!k| {W(k \!- \!1) \!=\! m\! - \!1,L(k \!-\! 1)\! = \!{L_h}\! -\! k \!+ \!1}\}$. As shown in Fig. \ref{Simplified}, in the case of $2\!\leq \!m\!<\! L_h$, the consensus process cannot go through the repeated confirmation path according to the definition of confirmation delay. Hence, the first line in (\ref{H2LR_d}) is to ensure that the observed transaction reaches confirmation though the valid paths in Fig. \ref{Simplified}. In the case of $m\geq L_h$, \{$W(k)=m, L(k)=1$\} is the only state for confirmation.

\textbf{L2HR:} In this regime, the cumulative weight of an observed transaction increases by $1$ with probability $1$ in each transaction arrival interval. The expected confirmation delay $E[T_{l2hr}]$ in L2HR can be expressed as follows.
\begin{equation}\label{L2HR_d}
{E[T_{l2hr}]} \!=\!E\sum\limits_{i=1}^{m-1} {{H_i}}\!= \!{(m\!-\!1)}{h_{h}},~~~~m\!\geq\!2,
\end{equation}
where $h_{h}\!=\!1/\lambda_{h}$.

\section{Double-spending Attack Model}

In this section, we first introduce the most typical double-spending attack in a DAG-based ledger. Then, we use a stochastic model to examine the probability of a successful double-spending attack.

\subsection{Attack Descriptions and Assumptions}

In preliminaries, we mentioned that a DAG-based ledger uses the cumulative computational power of honest nodes to prevent data from tampering, and meanwhile, the cumulative computational power is proportional to cumulative weight. When the transaction arrival rate is low, the cumulative weight growth rate will decrease, and it would be easy for an attacker to outweigh the cumulative weight of the branch maintained by the honest nodes for double-spending. Moreover, as we analysed before, the consensus process is affected by network load. Therefore, a rational attacker would optimise its strategy by considering network load to increase the success probability.

\begin{figure}[t]
\setlength{\abovecaptionskip}{0.cm}
\setlength{\belowcaptionskip}{-0.cm}
\captionsetup{font={footnotesize}}
\begin{center}
\includegraphics[width=8cm]{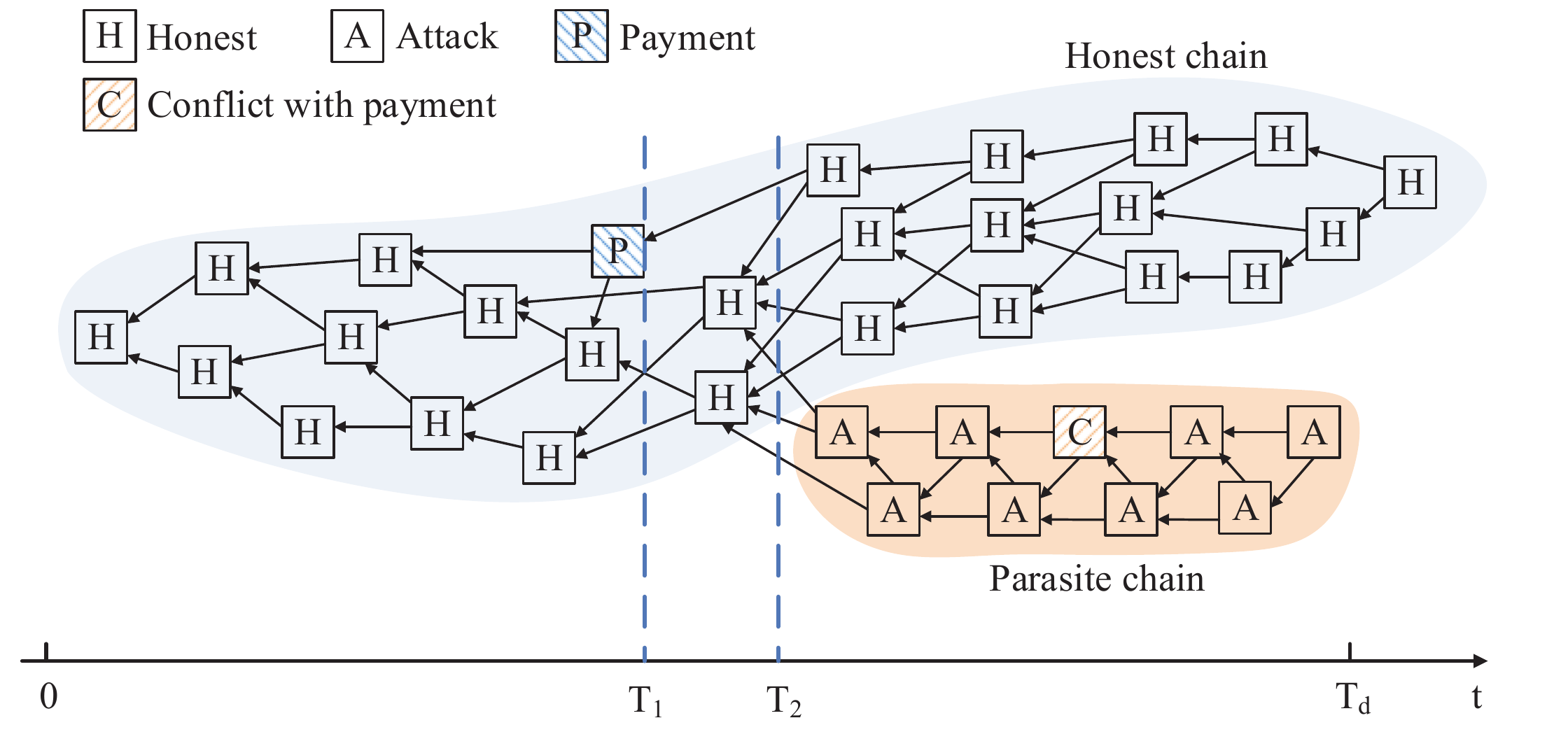}
\end{center}
\caption{Parasite chain for double-spending attack}
\label{Parasite}
\end{figure}

To systematically analyse this problem, we introduce the most typical double-spending attack in the DAG-based ledger, the parasite chain attack, which is shown in Fig. \ref{Parasite}. \begin{enumerate}
  \item Let $T_{1}$ be the time when the attacker sends a payment to a merchant and the honest nodes begin to approve it.
  \item Let $T_{2}$ be the time when the attacker builds an offline branch (called as parasite chain) and no honest node knows that. The parasite chain contains a transaction that conflicts with the payment. Note that this could be acted before $T_1$. In other words, $T_1<T_2$ or $T_1\geq T_2$ are both allowed (we will analyse these two cases later).
  \item The attacker continually uses its computational power to perform hash operations, and issues new transactions to extend the parasite chain for increasing its overall cumulative weight.
  \item Let $T_{d}$ be the time when the payment for merchant reaches confirmation threshold $m$, so the merchant sends goods to the attacker.
  \item As long as the cumulative weight of the parasite chain outweighs the honest chain after $T_{d}$, the attacker will broadcast the parasite chain to the whole network. The honest nodes will select the parasite chain gradually based on MCMC algorithm. The payment for merchant will be orphaned finally, but the goods (e.g., a piece of useful message) have already been sent to the attacker, so the double-spending attack is successful.
\end{enumerate}

Next, we present the assumptions for double-spending analysis. Assume that the process of incoming new transactions issued by honest nodes follows a Poisson process with $\lambda$. Assume that the time of an attacker to perform hash operations to meet the targets\footnote{The targets are the hash value which begin with a specified number of zero bits announced by system.} is exponentially distributed having mean $1/\mu$ \cite{3-Mastering}.

\emph{Proof:} according to the widely used Keccak-384 hash algorithm \cite{3-SHA}, all results of this hash algorithm are in $(0,{\rm{ }}{{\rm{2}}^{384}}]$. As a result, the probability to meet the target is
\begin{equation}
\frac{the \ number \ of \ targets}{{2^{384}} - 1 - hashrate \times time}.
\end{equation}
Considering the current hashrate of mining pool is $45EH/s$ in Mar. 2019 \cite{BTC}, the practical progress of hash operations ($hashrate \times time$) is still much less than $2^{384}$. Meanwhile, since the hashrate of honest nodes and attackers in DAG network are much less than mining pool usually, the impact of hash operations progress on the probability to meet the target is negligible. This means the hash operations process in distributed ledgers can be treated as memoryless.

\subsection{Probability of A Successful Attack}

Based on the previous assumption of the own weight of each transaction is $1$, the attack would be successful when the number of transactions issued by the attacker are more than that by honest nodes after $T_{d}$.

We can divide the competition process between the attacker and honest nodes into multiple rounds. Each round depicts the overall number of issued transactions increasing by $1$. Suppose the attacker creates a parasite chain by extending tips at $T_{2}$. The competition begins and the overall number of issued transactions at two branches is $0$ at this moment.

Let $\{H_{i}, i\!=\! 1, 2, . . .\infty \}$ denote the sequence of interarrival times between two neighbouring transactions, where $H_{i}, i \!=\! 1, 2, . . .\infty,$ are independent identically distributed exponential random variables with mean $1/\lambda$. Let $\{A_{j}, j \!=\! 1, 2, . . .\infty \}$ be the sequence of interarrival times of transactions issued by the attacker, where $A_{j}, j\!=\! 1, 2, . . .\infty,$ are independent identically distributed exponential random variables with mean $1/\mu$.

In the first round, according to \cite{14-Probability}, we can obtain the probability that one exponential random variable is smaller than another as follows.
\begin{equation}\small
\begin{split}
&P\{the \ transaction \ in \ round \ 1 \ is \ issued \ by \ honest \ nodes\}\!\\
&=\!P\{ {H_1}\!<\!{A_1}\}\!=\!\frac{\lambda }{{\lambda\!+\!\mu }},
\end{split}
\end{equation}
\begin{equation}\small
\begin{split}
&P\{the \ transaction \ in \ round \ 1 \ is \ issued \ by \ the \ attacker \}\!\\
&=\!1\!-\!\frac{\lambda }{{\lambda\!+\!\mu }}\!=\!\frac{\mu }{{\lambda\!+\!\mu }}.
\end{split}
\end{equation}

\begin{figure*}[t]
\captionsetup{font={footnotesize}}
\begin{minipage}[t]{0.32\linewidth}
\centering
\includegraphics[width=2in, height=1.8in]{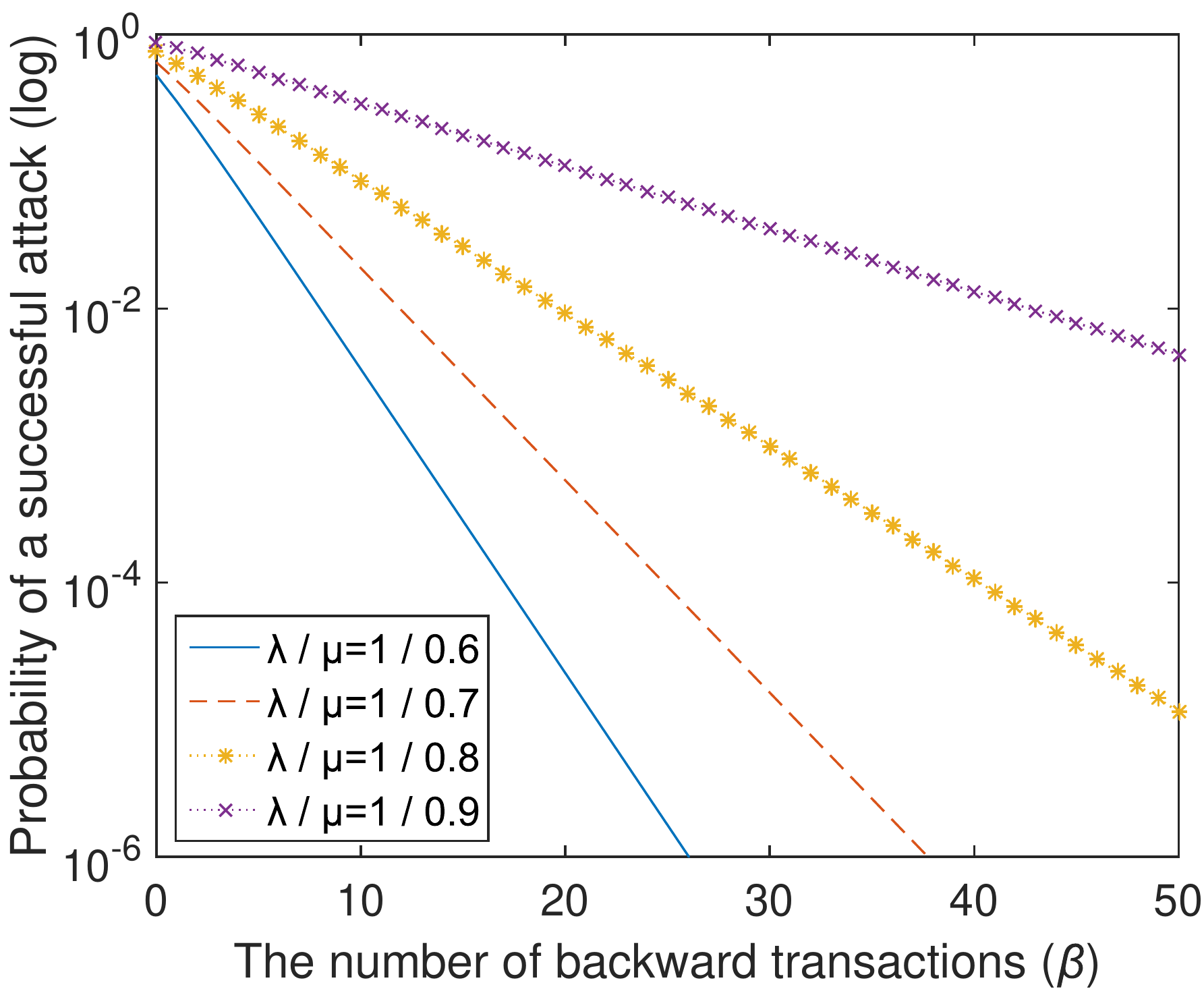}
\caption{Successful attack probability vs. $\beta$}
\label{PvsK}
\end{minipage}
\begin{minipage}[t]{0.32\linewidth}
\centering
\includegraphics[width=2in, height=1.8in]{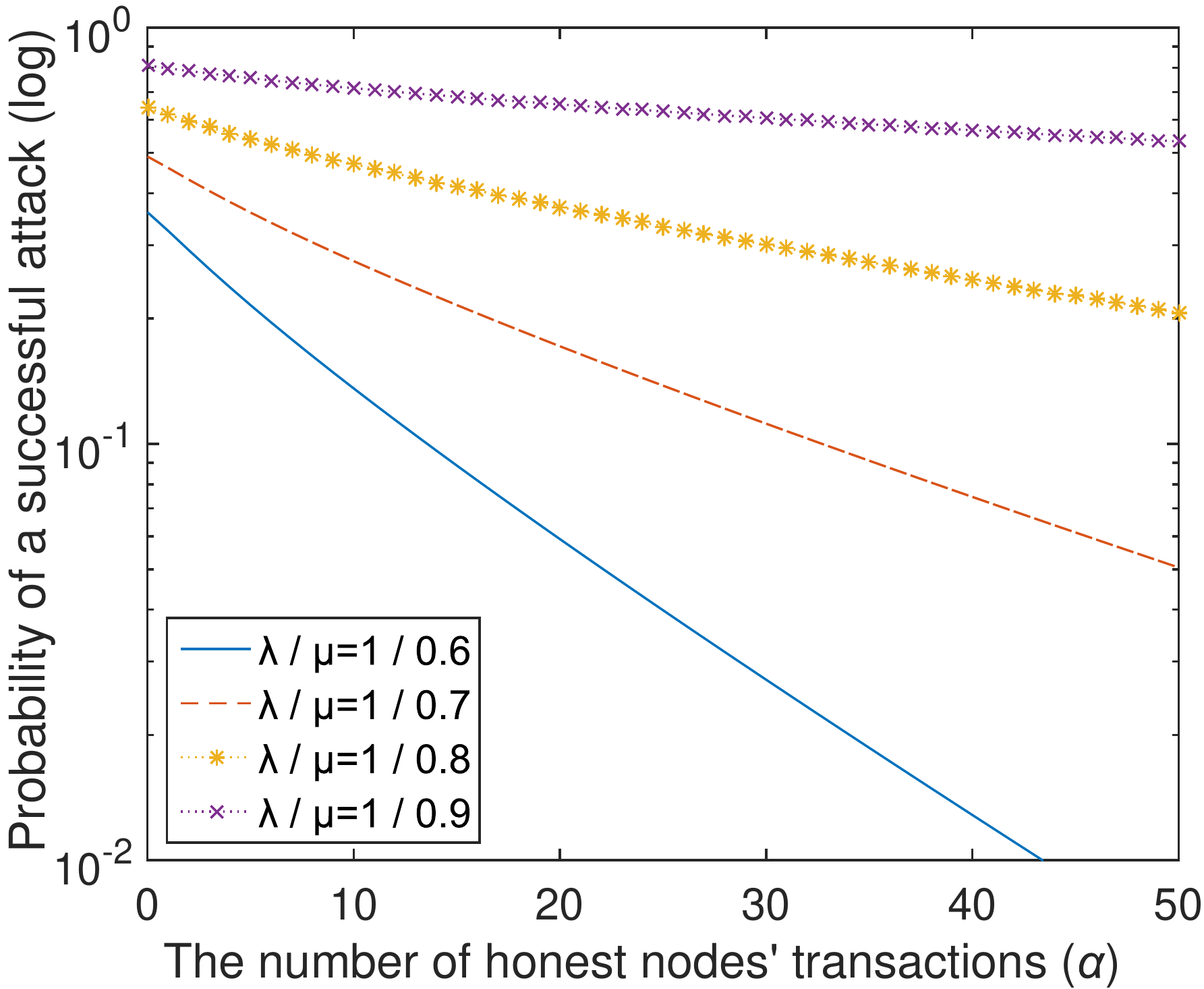}
\caption{Successful attack probability vs. $\alpha$}
\label{PvsR}
\end{minipage}
\begin{minipage}[t]{0.32\linewidth}
\centering
\includegraphics[width=5.5cm]{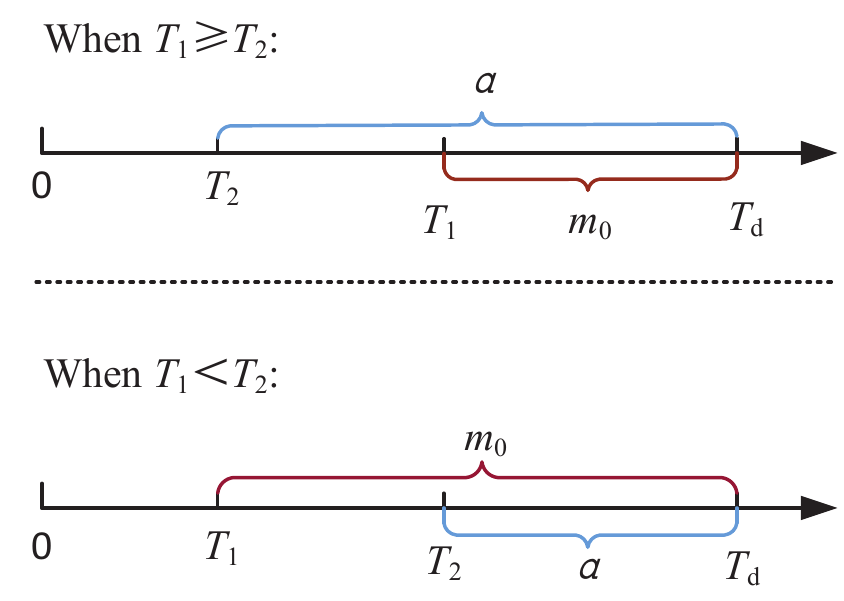}
\caption{The influence of $T_{1}$ and $T_{2}$ on $\alpha$}
\label{T1vsT2}
\end{minipage}
\end{figure*}

In the second round, if the first transaction is issued by honest nodes, we have
\begin{equation}\small
\begin{split}
&P\{the \ transaction \ in \ round \ 2 \ is \ issued \ by \ honest \ nodes\}\!\\
&=\!P\{ {H_2} \!<\! {A_1}\! -\! {H_1}\left| {{H_1}\! < \!{A_1}} \right.\}\!=\! P\{ {A_1}\! >\! {H_2}\! + \!{H_1}\left| {{A_1} \!> \!} \right.{H_1}\}\\
&=\! P\{ {A_1}\! >\! {H_2}\} \! = \!\frac{\lambda }{{\lambda  \!+ \!\mu }}.
\end{split}
\end{equation}

Alternatively, if the first transaction is issued by the attacker, we have
\begin{equation}\small
\begin{split}
&P\{the \ transaction \ in \ round \ 2 \ is \ issued \ by \ honest \ nodes\}\!\\
&=\!1\!-\!P\{ {H_1}\! -\! {A_1} \!>\! {A_2}\left| {{H_1}\! > \!{A_1}} \right.\}\!=\! 1\!-\!P\{ {H_1}\! >\! {A_1}\! + \!{A_2}\left| {{H_1} \!> \!} \right.{A_1}\}\\
&=\! 1\!-\!P\{ {H_1}\! >\! {A_2}\} \! = \!\frac{\lambda }{{\lambda  \!+ \!\mu }}.
\end{split}
\end{equation}

Generally, in any round, we have
\begin{equation}\label{p}\small
\begin{split}
&P\{the \ new \ transaction \ is \ issued \ by \ honest \ nodes\}\!\\
&=\!P\{ {H_i}\!<\!{A_j}\}\!=\!\frac{\lambda }{{\lambda\!+\!\mu }},
\end{split}
\end{equation}
\begin{equation}\label{q}\small
P\{the \ new \ transaction \ is \ issued \ by \ the \ attacker \}\!=\!\frac{\mu }{{\lambda\!+\!\mu }}.
\end{equation}

Let the probability in (\ref{p}) be $p$ and that in (\ref{q}) be $q$, the attack process can be treated as independent Bernoulli trials.

Accordingly, we analyse the attack process before $T_{d}$. In this process, the attacker cannot broadcast its parasite chain even if it outweighs the honest chain at some point, since the merchant has not sent goods yet. Let $\alpha$ be the number of transactions issued by honest nodes from $T_{2}$ to $T_{d}$, and $N$ be the possible number of transactions issued by the attacker when the honest nodes have issued $\alpha$ transactions. Based on negative binomial distribution theory \cite{8-double-spend}, the probability mass function of $N$ can be given as
\begin{equation}\label{N}
P\{ N \!=\! n\}  \!=\! \dbinom{n \!+\! \alpha\! -\! 1}{\alpha \!-\! 1}{p^\alpha}{q^n},~~~\alpha\!\geq1.
\end{equation}

If $N\!>\!\alpha$, the parasite chain attack will succeed at $T_{d}$. Otherwise, in order to win, the attacker should catch up the difference of issued transactions until the parasite chain outweighs the honest chain after $T_{d}$. This event is analogous to a Gambler's Ruin problem \cite{8-double-spend}, the attacker should catch up the difference of $\alpha \!- \!N \!+ \!1$ transactions at least, and the corresponding probability to catch up is shown as follows,
\begin{equation}
{P_c}(\alpha\! -\! N \!+\! 1)\! =\! \left\{ {\begin{array}{*{20}{l}}
{{{(q/p)}^{\alpha-N+1}},  \ \ \ \    p \! > \! q\ {\rm{and}} \ \alpha\!\geq1},\\
{1,   \ \ \ \ \ \ \ \ \ \, \, \ \ \ ~~~~    p\! \leq \!q}.
\end{array}} \right.
\end{equation}

In summary, the probability of a successful double-spending attack when $\alpha\!\geq1$ is
\begin{equation}\small\label{23}
\begin{split}
&P\{ attack \ succeeds\} \! = \!P\{ N\! >\! \alpha\} \! +\! P\{ N \le \alpha\} {P_c}(\alpha\! -\! N\! +\! 1) \\
&= \sum\limits_{n = \alpha \!+\! 1}^\infty
\dbinom{n \!+\! \alpha\! -\! 1}{\alpha \!-\! 1}{p^\alpha}{q^n} \! + \!\sum\limits_{n=0}^\alpha \dbinom{n \!+\! \alpha\! -\! 1}{\alpha \!-\! 1}{p^\alpha}{q^n}{{(\min(q/p,1))}^{\alpha\! -\! n\! +\! 1}} \\
&= \left\{ {\begin{array}{*{20}{l}}
{1 \!-\! \sum\limits_{n=0}^\alpha \dbinom{n \!+\! \alpha\! -\! 1}{\alpha \!-\! 1}({p^\alpha}{q^n}\!- \!{p^{n\! - \!1}}{q^{\alpha\! + \!1}} ) ,  \ \ \ \ \   p \! > \! q \ {\rm{and}} \ \alpha\!\geq1},\\
{1, ~~~~~~~~~~~~~~~~~~~~~~~~~~~~~~~~~~~~~~~~~~~~~~~~~ p\! \leq \! q}.
\end{array}} \right.
\end{split}
\end{equation}

Especially, when $\alpha\!=\!0$, the attacker should build the parasite chain as soon as the honest payment is confirmed, and in this case we can have $T_{2}\!=\!T_{d}$. As a result, the competition before $T_{d}$ disappears. However, in order to outweigh honest chain, the attacker also should outpace honest nodes by $1$ transaction at least after $T_{d}$. The probability of a successful attack in the case of $\alpha=0$ is
\begin{equation}\label{24}\small
P\{ attack \ succeeds\} \! = \!{P_c}(1)\! =\! \left\{ {\begin{array}{*{20}{l}}
{q/p, ~~~   p \! > \! q\ {\rm{and}} \ \alpha\!=\!0},\\
{1, ~~~~~ \,    p\! \leq \!q}.
\end{array}} \right.
\end{equation}

By integrating (\ref{23}) and (\ref{24}), the probability of a successful double-spending attack is
\begin{equation}\small\label{SP}
\begin{split}
&P\{ attack \ succeeds\} \\
&= \left\{ {\begin{array}{*{20}{l}}
{q/p, ~~~~~~~~~~~~~~~~~~~~~~~~~~~~~~~~~~~~~~~~~~~ \,  p \! > \! q\ {\rm{and}} \ \alpha\!=\!0},\\
{1 \!-\! \sum\limits_{n=0}^\alpha \dbinom{n \!+\! \alpha\! -\! 1}{\alpha \!-\! 1}( {p^\alpha}{q^n}\!- \!{p^{n\! - \!1}}{q^{\alpha\! + \!1}}) ,  \  \   p \! > \! q \ {\rm{and}} \ \alpha\!\geq1},\\
{1, ~~~~~~~~~~~~~~~~~~~~~~~~~~~~~~~~~~~~~~~~~~~~~~ p\! \leq \! q},
\end{array}} \right.
\end{split}
\end{equation}
where $p \!=\! \lambda/(\lambda\!+\!\mu)$, $q \!=\! \mu/(\lambda\!+\!\mu)$.

In this work, we use $\lambda$ and $\mu$ representing the transaction arrival rate of honest nodes and the attacker to conduct the double-spending analysis. To extend this attack model to other consensus algorithm such as PoS, we only need to change is the way to generate $\lambda$ and $\mu$, e.g., using balance (stake in PoS) to replace computational power.


\section{Security Analysis}

In this section, we analyse the strategy to increase the probability of a successful parasite chain attack on the perspective of the attacker. Based on (\ref{SP}), the probability of a successful attack is identically equal to $1$ when $p\!\leq \!q$ (i.e., $\lambda\!\leq\!\mu$). So we only analyse the situation when $p\!>\!q$. 

\subsection{Attack Strategy}

\subsubsection {How to attach the parasite chain into DAG}

If the attacker builds a parasite chain on earlier transactions that have been approved by some other transactions at $T_{2}$, it needs to catch up the difference between the honest chain and its own from the start, which is generated by the number of transactions from the selected earlier transactions to tips. Let the difference be $\beta$, at $T_d$, the attacker should issue $\alpha\!+\!\beta\!+\!1$ at least to succeed. Otherwise, after $T_d$, the attacker should catch up the difference of $\alpha+\!\beta\!-\!N\!+\!1$ transactions. The corresponding probability is
\begin{equation}
{P_c}(\alpha \!+ \!\beta \!-\! N\! +\! 1)\! =\! {(q/p)^{\alpha + \beta - N + 1}},~~p \! > \! q \ {\rm{and}} \ \alpha\!\geq\!1.
\end{equation}

Especially, when $\alpha\!=\!0$, the attacker should catch up the difference of $\beta+1$ transactions at least after $T_d$. The probability of a successful attack for $\alpha\!=\!0$ is ${P_c}(~\!\beta\! +\! 1)$.

In summary, when the attacker builds a parasite chain on earlier transactions, the probability of a successful attack is
\begin{equation}\small\label{beta}
\begin{split}
&P\{attack \ succeeds \ with \ the\ difference \ \beta\} \! \\
&= \left\{ {\begin{array}{*{20}{l}}
{{P_c}(~\!\beta\! +\! 1),~~~~~~~~~~~~~~~~~~~~~~~~~~~~~~~~~~~~~~~~~ \, ~~~~~~~~~~~\alpha\!=\!0}\\
{P\{ N\! >\! \alpha\! + \!\beta\} \! +\! P\{ N \!\le\! \alpha \!+\! \beta\}{P_c}(\alpha \!+\! \beta\! -\! N \!+\! 1),~~~~~~ \, \alpha\!\geq\!1}
\end{array}} \right.\\
&= \left\{ {\begin{array}{*{20}{l}}
{(q/p)^{\beta+ 1}, ~~~~~~~~~~~~~~~~~~~~~~~~~~~~~~~~~~~~~~~~p \! > \! q \ {\rm{and}} \ \alpha\!=\!0},\\
{\!1\! -\! \sum\limits_{n = 0}^{\alpha \!+\! \beta} \dbinom{n \!+\! \alpha \!- \!1}{\alpha\! -\! 1}({p^\alpha}{q^n} \!-\!{p^{n\!-\!\beta\!-1}}{q^{\alpha\!+\!\beta\!+\!1}}) ,   \     p \! > \! q \ {\rm{and}} \ \alpha\!\geq\!1}.
\end{array}} \right.
\end{split}
\end{equation}


To capture the impact of $\beta$, we use (\ref{beta}) to conduct a case study and let $\alpha\!=\!1$. The results in Fig. \ref{PvsK} clearly illustrate that the probability of a successful attack decreases with $\beta$, which shows the impact of $\beta$ on the attack. Moreover, we can see that $\beta$ is generated when the attacker does not choose tips to build the parasite chain. As a result, it is a natural option to choose tips for the attacker if possible, which can increase the probability of a successful attack with the minimum $\beta$.

\subsubsection {Minimize the number of transactions of honest chain from $T_{2}$ to $T_{d}$}



Intuitively, when $p\!>\!q$, the transaction arrival rate on the honest chain is higher than that of parasite chain, and thus the probability of a successful attack would be declined with the increasing of $\alpha$ on the honest chain from $T_{2}$ to $T_{d}$. Different from the previous case that shows the impact of $\beta$, we conduct another case study to investigate the impact of $\alpha$ using (\ref{beta}), where $\beta=1$.


In Fig. \ref{PvsR}, we can see that the probability of a successful attack declines obviously with the increasing of $\alpha$. The reason is that the larger $\alpha$ indicates the higher cumulative weight of honest chain and it would be safer. As a result, the attacker should invest much more computational power against the larger $\alpha$, otherwise, it is difficult to succeed.


Therefore, the attacker should also minimize $\alpha$ to optimise its attack strategy. Moreover, we know that $\alpha$ is determined by the time in attack process shown in Fig. \ref{T1vsT2}, and thus the attacker can adjust its action at the right time to minimize $\alpha$ as follows. Denote the number of transactions issued by honest nodes from $T_{1}$ to $T_{d}$ as $m_{0}$, it is a constant value for a specific attack. As shown in Fig. \ref{T1vsT2}, in order to decrease $\alpha$, we can see that the duration between $T_1$ and $T_2$ is the less the better when $T_1\!\geq \!T_2$. In contrast, it is the more the better when $T_1\!<\! T_2$.

However, the attacker cannot defer $T_{2}$ indefinitely for decreasing $\alpha$. By comparing Fig. \ref{PvsK} (the lowest value shown is $10^{-6}$) with Fig. \ref{PvsR} (the lowest value shown is $10^{-2}$), we could notice that the decline rate of probability in Fig. \ref{PvsK} is faster than that in Fig. \ref{PvsR}, which reflects the impact of $\beta$ is higher than $\alpha$. Therefore, to maximize success probability, the attacker should first follow the strategy of building the parasite chain on tips to minimize $\beta$, then postpone $T_{2}$ to the time before the honest payment has been indirectly approved by all the tips. Since if $T_{2}$ is later than that time, the parasite chain for double-spending will indirectly approve the honest payment, and the attack cannot succeed.

In summary, to launch a better parasite chain attack, the attacker should minimize $\alpha$ and $\beta$ by choosing the tips to build a parasite chain at the last time before the honest payment has been indirectly approved by all tips.



\subsection{Adopt Attack Strategy in Different Load Regimes}

Next, we analyse how to determine the strategy to increase the probability of a successful attack according to the network load. To distinguish the impact of network load on $p$ and $q$, let $p_{h}\!=\!\lambda_{h}/(\lambda_{h}\!+\!\mu)$, $q_{h}\!=\!\mu/(\lambda_{h}\!+\!\mu)$ in HR and $p_{l}\!=\!\lambda_{l}/(\lambda_{l}\!+\!\mu)$, $q_{l}\!=\!\mu/(\lambda_{l}\!+\!\mu)$ in LR, respectively.

\textbf{HR:} 
According to the physics meaning of adaptation period in Section IV, the attacker should build the parasite chain at the end of adaptation period, which is the best time for $T_2$. At this moment, the honest payment will be indirectly approved by all tips very soon, and the expected cumulative weight of the honest payment at $T_2$ is $W(t_0)\!-\!1$. Meanwhile, based on the definition of $\alpha$, we have $\alpha\!=\!\max \{m \!- \![W(t_{0})] \!+ \!1,0\}$. Let $f_{h}(x) \!= \! 1 \!-\! \sum\limits_{n = 0}^x \dbinom{n \!+\! x \!- \!1}{x \!-\! 1}({p_{h}^x}{q_{h}^n}\!- \!{p_{h}^{n - 1}}{q_{h}^{x + 1}} )$, we can obtain the probability of a successful attack in HR based on (\ref{SP}), which is expressed as follows.
\begin{equation}\label{SPHR}
\begin{split}
&P\{attack \ succeeds \ in \ HR\} \! \\
&= \! \left\{ {\begin{array}{*{20}{l}}
{q_{h}/p_{h}, \ \ \ \ \ \ \ \ \ \  \ ~~~~~ ~ p_{h} \! > \! q_{h} \ {\rm{and}} \ 2 \!\le \! m \! < \! [W(t_{0})]},\\
{f_{h}(m \!- \![W(t_0)] \!+\! 1),     \ p_{h}  \!> \! q_{h} \  {\rm{and}} \ m\! \geq \! [W(t_{0})]},\\
{1,  \ \ \  \ \ \ \ \ \ \ \ \ \ \ \ \ ~ \ \ \ \ \ \,      p_{h} \! \le \! q_{h}}.
\end{array}} \right.
\end{split}
\end{equation}

\begin{figure*}[t]
\captionsetup{font={footnotesize}}
\begin{minipage}[t]{0.32\linewidth}
\centering
\includegraphics[width=2in, height=1.8in]{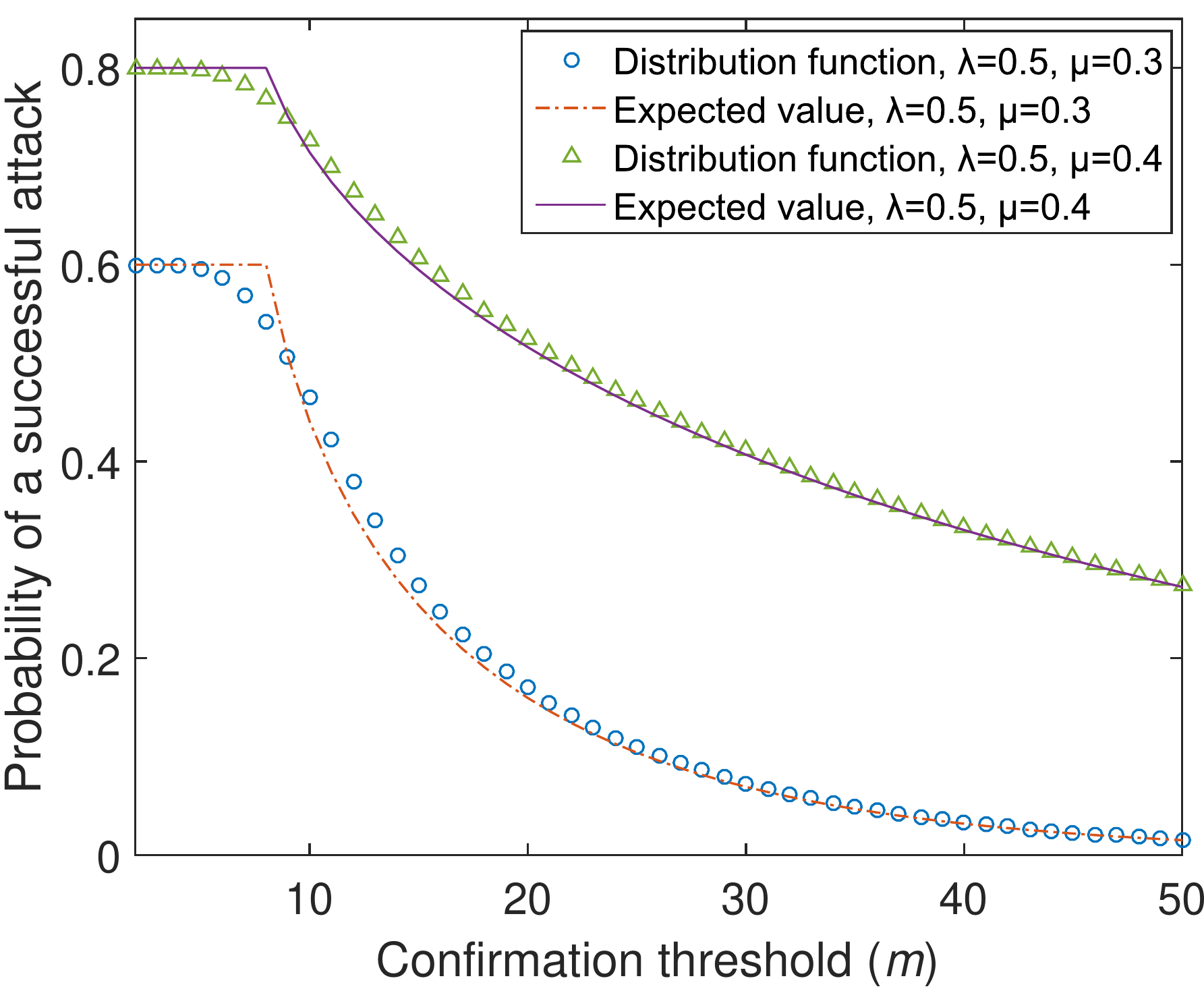}
\caption{Probability of a successful attack with expected value and distribution function}
\label{17}
\end{minipage}
\begin{minipage}[t]{0.32\linewidth}
\centering
\includegraphics[width=2in, height=1.8in]{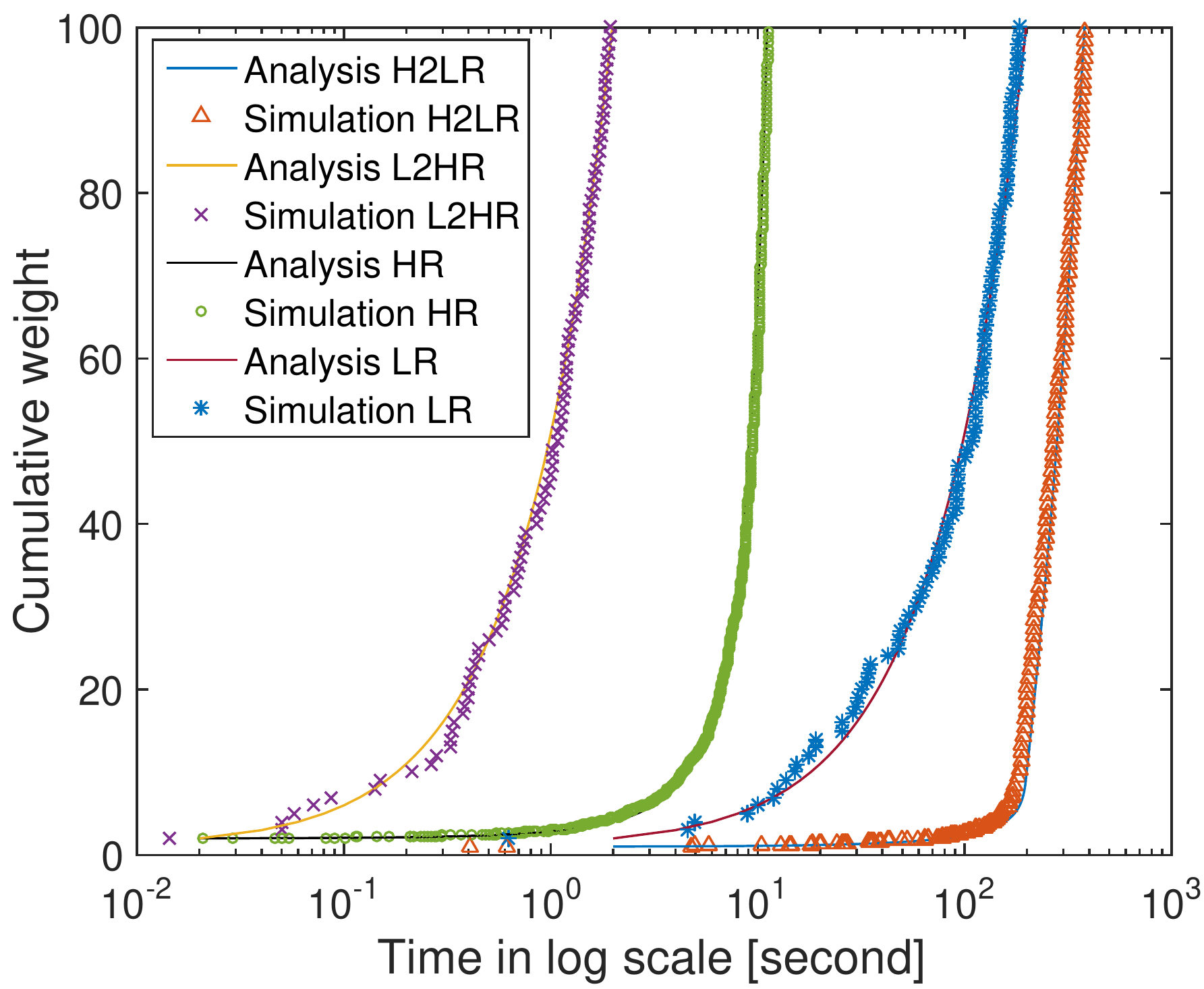}
\caption{Cumulative weight vs. time}
\label{CW}
\end{minipage}
\begin{minipage}[t]{0.32\linewidth}
\centering
\includegraphics[width=2in, height=1.8in]{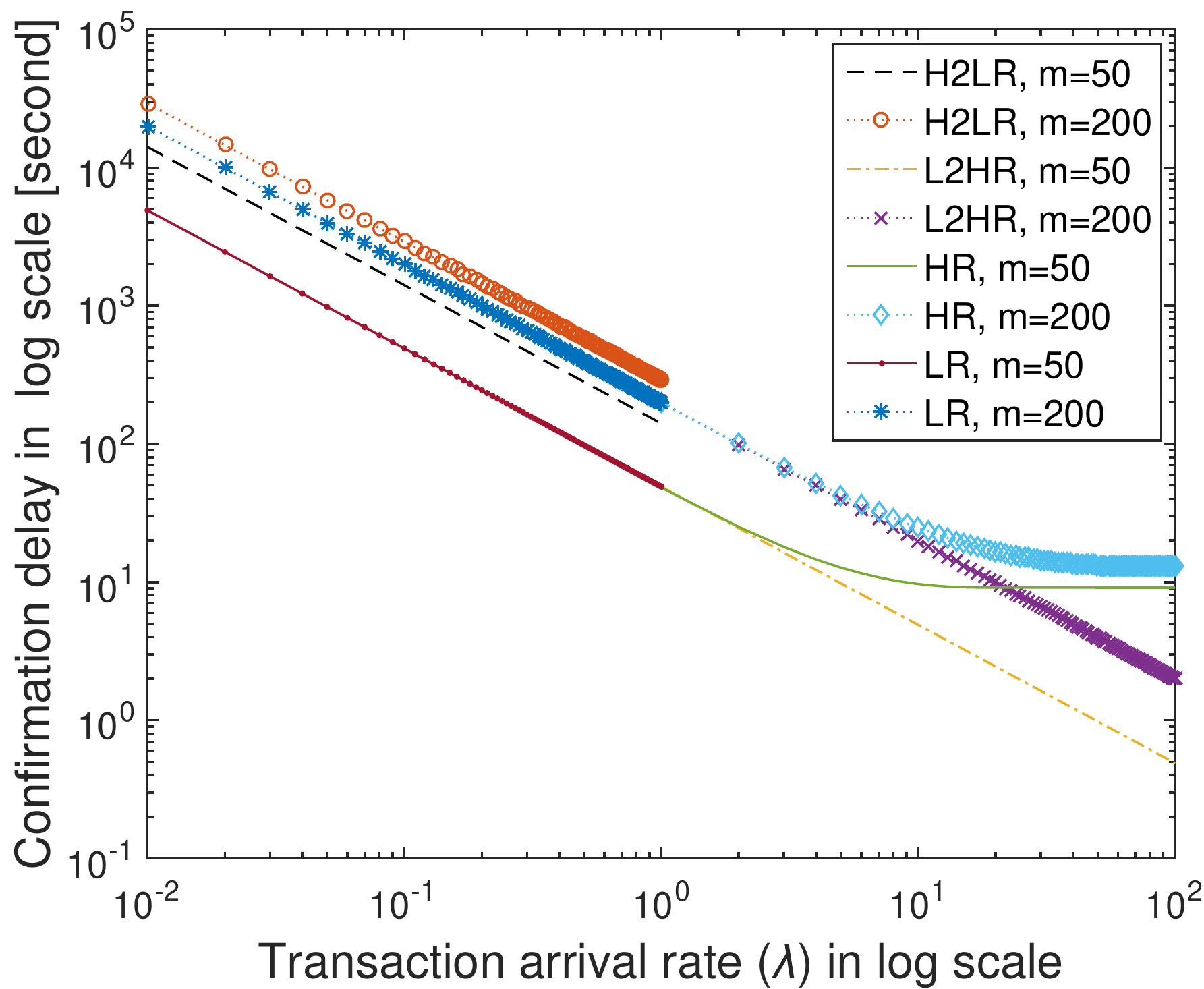}
\caption{Confirmation delay vs. $\lambda$}
\label{CD}
\end{minipage}
\end{figure*}


\textbf{LR:} As mentioned before, the DAG-based ledger can be treated as a single chain since $L(0)\!=\!1$ in this regime. The honest payment is indirectly approved by all tips at $T_1$. According to the analysis of attack strategy, we can know the best $T_2$ in LR is $T_2\!=\!T_1$. However, since the honest payment is the only tip as soon as it reveals, the attacker can only attach the parasite chain before it, and thus the best case is $\beta\!=\!1$. Meanwhile, since the own weight of honest payment is $1$, we can obtain that $\alpha\!=\!m\!-\!1$. Based on $m\!\geq\!2$, we have $\alpha\! \geq\!1$. Using (\ref{beta}), the probability of a successful attack in LR is
\begin{equation}\label{SPLR}
\begin{array}{l}
P\{ attack \ succeeds \ in \ LR \} \\
 = \left\{ {\begin{array}{*{20}{c}}
{1 \!-\! \sum\limits_{n = 0}^{m} \dbinom{n \!+\! m \!- \!2}{m \!-\! 2}({p_{l}^{m\!-\!1}}{q_{l}^n} \!- \!{p_{l}^{n \!-\! 2}}{q_{l}^{m \!+ \!1}}) , ~ \,   p_{l}  \!> \! q_{l}},\\
{1,~~~~~~~~~~~~~~~~~~~~~~~~~~~~~~~~~~~~~~~~~~~~~~~ \, p_{l} \!\le \!q_{l}}.
\end{array}} \right.
\end{array}
\end{equation}

\textbf{H2LR:} In this regime, the number of tips would decrease from $L(0)\!=\!L_{h}\!=\!2\lambda_{h}h_r$ to $L(k)\!=\!1$ finally.
The honest payment will be indirectly approved by all tips when the number of tips becomes $2$, and the attacker should build the parasite chain at this moment. According to the Markov chain in Fig. 4, we can obtain the possible states of the honest payment at $T_{2}$ is $\{W({L_h} \!-\! 2) \!=\! i,L({L_h} \!-\! 2) \!= \!2\}$, where $i \!=\! 1,2, \cdots ,L_{h}\!- \!1$. Accordingly, after $T_{2}$, the honest payment needs $\max \{m \!- \!i,0\}$ approvals at least to reach confirmation threshold $m$, which means $\alpha\!=\!\max \{m \!- \!i,0\}$. Based on (\ref{SP}), the probability of a successful attack in H2LR can be expressed as
\begin{equation}\small\label{SPH2LR}
\begin{split}
&P\{attack \ succeeds \ in \ H2LR\} \! \\
&= \! \left\{ {\begin{array}{*{20}{l}}
{\sum\limits_{i  =  1}^{m \! -\! 1} {{P_{\{i,2\}}}} \! \times \! f_{l}(m \!-\! i) \! +\!  \sum\limits_{i  =  m}^{{L_h} \! - \! 1} {{P_{\{i,2\}}}} \! \times  \! {q_{l}/p_{l}}, \ p_{l} \! > \! q_{l}, 2 \!\le \! m \! < \! L_{h}},\\
{\sum\limits_{i = 1}^{{L_h} \!-\! 1} {{P_{\{i,2\}}}} \! \times \! f_{l}(m \!- \!i), ~~~~~~~~~~~~~~~~~~~~~~~ \, \ p_{l}  \!> \! q_{l}, m \! \geq \! L_h },\\
{1  ~~~~~~~~~~~~~~~~~~~~~~~~~~~~~~~~~~~~~~~~~~~~~~~~~~ \,    p_{l} \! \le \! q_{l}},
\end{array}} \right.
\end{split}
\end{equation}
where $f_{l}(x) \!=\! 1 \!-\! \sum\limits_{n = 0}^x \dbinom{n \!+\! x \!- \!1}{x \!-\! 1}({p_{l}^x}{q_{l}^n} \!- \!{p_{l}^{n - 1}}{q_{l}^{x + 1}})$ and {\small${P_{\{i,2\}}} \!=\! P\left\{ {W({L_h} \!-\! 2) \!=\! i,L({L_h} \!-\! 2) \!=\! 2\left| {W(0) \!= \!1,L(0) \! =\! {L_h}} \right.} \right\}$, $i \!=\! 1,2, \cdots ,L_{h} - 1$.}

Note that it is very difficult to capture the probability distribution function of cumulative weight in HR due to too many possible states of covered transactions in this regime. Therefore, we use the expected value $W(t_{0})$ to evaluate the probability of a successful attack in HR. In contrast, since the distribution function of cumulative weight in H2LR can be calculated from Fig. 4, we have used it to analyse the probability of a successful attack shown in (\ref{SPH2LR}), which is different from HR.

In order to show the accuracy of the analysis using expected value, we conduct a case study to compare the analytical results based on expected value and distribution function in H2LR. Compared with (\ref{SPH2LR}) that is based on distribution function, the probability of a successful attack based on expected value in H2LR is
\begin{equation}\label{SPH2LR2}
\begin{split}
&P\{attack \ succeeds \  using \ expected \ value\} \! \\
&= \! \left\{ {\begin{array}{*{20}{l}}
{q_{l}/p_{l}, \ \ \ \ ~~~~~~~~~~~~ \ p_{l} \! > \! q_{l} \ {\rm{and}} \ 2 \!\le \! m \! < \! W_{0}},\\
{f_{l}(m \!- \!W_{0}),  \, \, \, ~~~~~~~~  p_{l}  \!> \! q_{l} \ {\rm{and}} \  m \! \geq \! W_0},\\
{1  \ \ \  \ \ \ \ \,  ~~~ ~~~~~~~~~~~~        p_{l} \! \le \! q_{l}},
\end{array}} \right.
\end{split}
\end{equation}
where $f_{l}(x) \!=\! 1 \!-\! \sum\limits_{n = 0}^x \dbinom{n \!+\! x \!- \!1}{x \!-\! 1}({p_{l}^x}{q_{l}^n} \!- \!{p_{l}^{n -1}}{q_{l}^{x + 1}})$, and the average cumulative weight of honest payment at the end of adaptation period in H2LR is ${W_0}\! =\! \sum\limits_{i = 1}^{L_{h} - 1} {{P_{\{i,2\}}}} \! \times \! i$, where {\small{${P_{\{i,2\}}} \!=\! P\left\{ {W({L_h} \!-\! 2) \!=\! i,L({L_h} \!-\! 2) \!=\! 2\left| {W(0) \!= \!1,L(0) \! =\! {L_h}} \right.} \right\}$, $i \!=\! 1,2, \cdots ,L_{h} - 1$.}}

To show the accuracy of the analysis using expected value, we use (\ref{SPH2LR}) and (\ref{SPH2LR2}) to conduct a comparison, where the initial number of tips $L_{h}$ is set as $100$. The result in Fig. \ref{17} shows that using expected value to evaluate the probability of a successful attack is feasible, and the probability evaluation results more accurate as long as the difference between $m$ and $W_{0}$ gets larger.



%
\textbf{L2HR:} Similar with LR, the DAG-based ledger can be also treated as a single chain in this regime, since the honest payment is indirectly approved by all tips at $T_{1}$. The expression of the probability to conduct a successful attack in this regime is similar to (\ref{SPLR}) but with $p_{h}$ and $q_{h}$, which is shown as follows.
\begin{equation}\label{SPL2HR}
\begin{array}{l}
P\{ attack \ succeeds \ in \ L2HR \} \\
 = \left\{ {\begin{array}{*{20}{c}}
{1 \!-\! \sum\limits_{n = 0}^{m} \dbinom{n \!+\! m \!- \!2}{m \!-\! 2}({p_{h}^{m\!-\!1}}{q_{h}^n} \!- \!{p_{h}^{n \!-\! 2}}{q_{h}^{m \!+ \!1}}) , ~ \,   p_{h}  \!> \! q_{h}},\\
{1,~~~~~~~~~~~~~~~~~~~~~~~~~~~~~~~~~~~~~~~~~~~~~~~ \, p_{h} \!\le \!q_{h}}.
\end{array}} \right.
\end{array}
\end{equation}


\section{Numerical Results and Discussions}

In this section, we numerically evaluate the performance of DAG consensus process in terms of cumulative weight and confirmation delay. Meanwhile, we show the probability of a successful double-spending attack in different load regimes.

The parameter settings are listed as follows. The transaction reveal delay is $h_{r}\!=\!1(s)$; the transaction arrival rate in HR is $\lambda_h\!=\!50$; the transaction arrival rate in LR is $\lambda_l\!=\!0.5$. Based on the analysis in Section III, the number of tips in HR and LR can be calculated as $L_{h} \!=\! 2\lambda_h h_{r} \!=\! 100$ and $L_{l} \!=\! 2\lambda_l h_{r} \!=\! 1$, respectively. To study the impact of confirmation threshold on the consensus process, we set $m \!=\! 50$, $100$ and $200$ to make a comparison. All the numerical results are obtained using Matlab.

\subsection{Cumulative Weight and Confirmation Delay}

In the first experiment, we calculate the cumulative weight growth of an observed transaction based on (\ref{HR_w}), (\ref{LR_w}), (\ref{H2LR_w}), and (\ref{L2HR_w}). To calculate the analytical results, we use the average interarrival time of new transaction ($1/\lambda_h$ or $1/\lambda_l$). Meanwhile, we use Poisson distribution to simulate the arrival process of new transactions and obtain simulation results.

%

Fig. \ref{CW} shows the growth trend of cumulative weight for an observed transaction under various load regimes. It is clear to see that the simulation results match well with their analytical results, which indicates the rightness and effectiveness of the proposed model. We can see that the cumulative weights for all load regimes increase over time, since the new transactions arrive continuously, and select tips to indirectly approve unconfirmed transactions. In this figure, HR and LR can act as a contrast to reflect the impact of network load as follows. Although the arrival rate $\lambda_{h}$ in L2HR and HR are the same, we can see that L2HR achieves $m$ faster than HR. This is because the initial number of tips in HR is much larger than that in L2HR, which results in a lower probability to select the observed transaction in HR, and thus a lower growth rate. 
Meanwhile, although the arrival rates are the same ($\lambda_l\!$) in LR and H2LR, we can see that LR outperforms H2LR. The reason is that adaptation period that is discussed in HR also exists in H2LR. In this period, the observed transaction has not been indirectly approved by all the tips, and thus the growth rate of cumulative weight in H2LR would be lower than $\lambda_l\!$ until the adaptation period ends. In contrast, the growth rate of cumulative weight in LR is $\lambda_l\!$ all the time, since without adaptation period, all new transactions will indirectly approve the observed transaction.



In the second experiment, using (\ref{H2LR_d}), (\ref{HR_d}), (\ref{LR_d}), and (\ref{L2HR_d}), we vary transaction arrival rate $\lambda$ to compare the confirmation delay under different load regimes.

Fig. \ref{CD} shows the confirmation delay of an observed transaction under various load regimes, we can see that $h_r\!=\!1$ can be seemed as a boundary between low and high network loads, where the performance of LR and H2LR are shown in $\lambda\!\in\![0,1]$, and the performance of HR and L2HR are shown in $\lambda\!\in\![1,100]$. The result demonstrates that the confirmation delay decreases with the increasing arrival rate. Meanwhile, for a given $m$, we can see the confirmation delay in H2LR is higher than LR and the confirmation delay in HR is higher than L2HR due to the impact of adaptation period, which matches well with the result in Fig. \ref{CW}. When $m$ changes from $50$ to $200$, the confirmation delay for all regimes increases. Meanwhile, the confirmation delay of unsteady regimes moves close to steady regimes due to a lower ratio of adaptation period to the whole consensus process. Moreover, we could notice that the confirmation delay in HR does not decrease linearly with increasing $\lambda$. This is because a higher $\lambda$ can result in a larger $W(t_{0})\!=\!\frac{2\lambda h_{r}}{0.704}$. Based on (\ref{HR_d}), when $m\!\leq\!W(t_{0})$, the observed transaction would be confirmed during adaptation period, and $\lambda$ plays no role in this case. The rationality behind this is the higher $\lambda$, the more number of tips based on $L\!=\!2\lambda h_{r}$, and thus the probability to select the observed transaction would decrease. So when $m\!\leq\!W(t_{0})$, even if the new transactions arrive faster, the confirmation delay would not decrease. Furthermore, with the increasing of $W(t_{0})\!=\!\frac{2\lambda h_{r}}{0.704}$, the curve of $m\!=\!50$ in HR reaches the lower bound of confirmation delay faster than that of $m\!=\!200$.

%
%

\subsection{Probability of A Successful Attack}

The following experiments are to examine the probability of a successful attack under steady and unsteady regimes based on network load.


\begin{figure}[t]
\setlength{\abovecaptionskip}{0.cm}
\setlength{\belowcaptionskip}{-0.3cm}
\captionsetup{font={footnotesize}}
\begin{center}
\includegraphics[width=5.5cm]{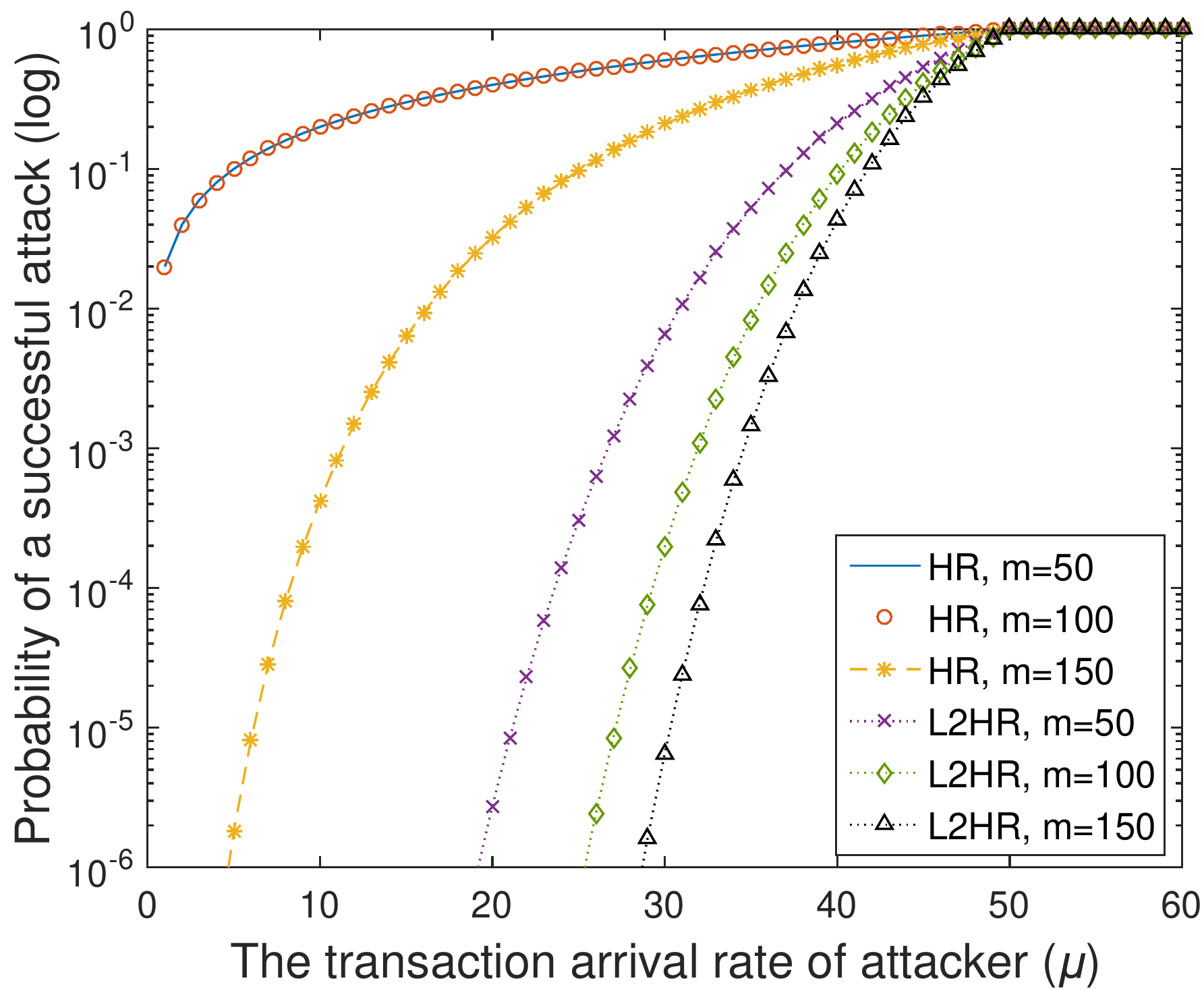}
\end{center}
\caption{The probability of a successful attack in HR and L2HR}
\label{HRvsL2HR}
\end{figure}

In the third experiment, using (\ref{SPHR}) and (\ref{SPL2HR}), we examine the probability of a successful attack in HR and L2HR by varying transaction arrival rate of the attacker $\mu$. Considering the confirmation threshold $m$ would result in different expressions in HR, which have been discussed in (\ref{SPHR}), we set $m=50,~100$ and $150$ based on the average cumulative weight at the end of adaptation period $W(t_{0})\!=\!\frac{L_{h}}{0.704}\!\approx\!142$. As a contrast, we also use the same $m$ in L2HR to illustrate the impact of adaptation period on the probability of a successful attack.

The result in Fig. \ref{HRvsL2HR} shows that when $\lambda\!>\!\mu$, the probability of a successful attack increases with $\mu$, and it is identically equal to $1$ when $\lambda\!\leq\!\mu$ based on the Gambler's Ruin problem. For a given confirmation threshold $m$, the probability of a successful attack in HR is higher than L2HR when $\lambda\!>\!\mu$. This is because HR has the adaptation period, and therefore, the attacker can ``steal" the computational power of the transactions that do not approve the honest payment by creating a parasite chain upon it.

Meanwhile, we notice that $m\!=\!50$ and $m\!=\!100$ in HR have the same success probability due to $W(t_{0})\!=\!142$. According to (\ref{SPHR}), as long as $m\!<\!W(t_{0})$, the honest payment would be confirmed during adaptation period, and thus $T_{2}\!=\!T_{d}$. The attacker only needs to outpace honest nodes by one transaction. Except that, we can find that a higher $m$ would result in a lower probability of a successful attack when the honest payment is confirmed during linear growth period. The reason is that the higher $m$, the more transactions issued by honest nodes from $T_{2}$ to $T_{d}$, and the harder for the attacker to outweigh honest chain since $\lambda\!>\!\mu$.




\begin{figure}[t]
\setlength{\abovecaptionskip}{0.cm}
\setlength{\belowcaptionskip}{-0.3cm}
\captionsetup{font={footnotesize}}
\begin{center}
\includegraphics[width=5.5cm]{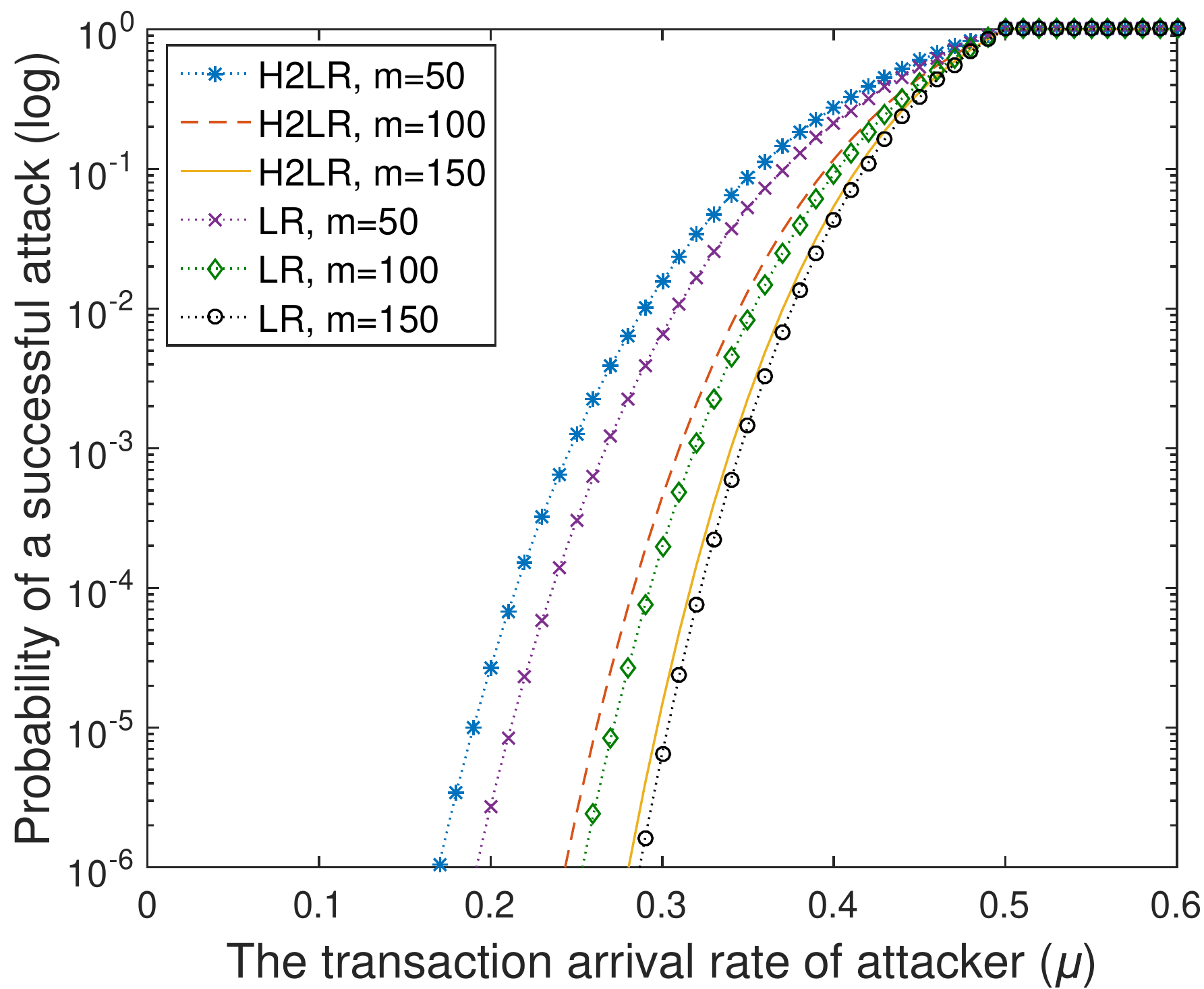}
\end{center}
\caption{The probability of a successful attack in LR and H2LR}
\label{H2LRvsLR}
\end{figure}

Finally, based on (\ref{SPLR}) and (\ref{SPH2LR}), we compare the probability of a successful attack in LR with that in H2LR. 
The result in Fig. \ref{H2LRvsLR} shows that the attacker could win with much less transaction arrival rate $\mu$ compare with that in Fig. \ref{HRvsL2HR}. This is because the transaction arrival rate of honest nodes is very low in H2LR and LR, thus, the cumulative weight of honest payment increases slowly. This phenomenon indicates that the low network load is harmful to the security. Meanwhile, Fig. \ref{H2LRvsLR} also reflects that the higher $m$, the success probability between H2LR and LR is closer. This means that a larger $m$ can result in a lower ratio of adaptation period to consensus process, which can reduce the adverse impact of adaptation period on security.

%
%
%
%

In summary, our analyses and experimental results reflect that the larger $m$ can reduce the adverse impact of adaptation period and decrease the probability of a successful attack in any regime. But on the other hand, a larger $m$ will result in a larger confirmation delay. Therefore, it is valuable to find a suitable confirmation threshold $m$ based on the trade-off between security level and confirmation delay according to the specific needs in a practical scenario.

\section{Related Work}

Besides DAG-based ledgers (e.g., Tangle \cite{9-tangle}, Byteball \cite{10-byteball} and Hashgraph \cite{11-hashgraph}), many other researches have been carried out to improve the throughput of traditional blockchain systems. Bitcoin-NG \cite{19-Bitcoin-NG} selects a leader to post multiple
blocks, thus increasing the block generation rate and the throughput. Hybrid-IoT \cite{20-Hybrid-IoT} proposes a two-tier blockchain architecture for IoT, where subgroups of IoT devices achieve
consensus through PoW algorithm and the connection among the sub-blockchains employs a Byzantine Fault-Tolerant (BFT) framework. Monoxide \cite{21-Monoxide} runs multiple independent and parallel PoW sub-blockchains termed as zones, in which different zones can conduct trading using the cross-zone algorithm. Although the high throughput can be achieved, the security in these systems are compromised since generating sub-blockchains will dilute the mining power of honest nodes. Meanwhile, due to the existence of single chain structure and PoW consensus algorithm, resource consumption and transaction fee are also the limitations of these distributed ledger technologies in IoT context. From this perspective, the DAG-based ledger is more appropriate for the IoT system, since it can satisfy high throughput, security and low cost simultaneously.

To validate this observation, mathematical models are required to quantitatively study the performance and limitation of different distributed ledgers. In \cite{20-Blockchain-enabled}, the authors analyse signalto-interference-plus-noise ratio, transaction transmission successful rate and throughput in blockchain-enabled IoT systems. Based on the performance analysis, the authors design an optimal node deployment algorithm for maximizing transaction throughput. In \cite{16-Stochastic}, the authors develop a stochastic model for the evolution and dynamics of blockchain networks, which provides a deeper understanding of crucial design issues for difficulty-of-work, block generation rate and adversarial
attacks. The above mathematical models is designed for PoW algorithm and the single chain structure. For DAG consensus, \cite{16-stability} examines the expected number of tips by formulating the tips selection algorithm as a ``balls into bins" problem. However, the balls into bins method is not precise enough,
since it conflicts with the fact that a new transaction cannot select one tip twice. As another work for DAG, \cite{16-Quasi-analytic} discusses the parasite chain absorption probabilities in the DAG-based ledger using two-way Markov chain Model. The author in this work focuses on the impact of cumulative weight on the result of the MCMC algorithm and does not study the double-spending attack.

Double-spending attack analysis is critical to a distributed ledger system. In \cite{1-bitcoin}, the author studies the require computational power for launching a double-spending attack in Bitcoin system by using Poisson distribution and Gambler's Ruin problem. However, due to the time for the honest nodes to find six blocks is variable, the Poisson distribution method relying on a constant time is not efficient enough. To improve this method, \cite{8-double-spend} uses negative binomial distribution to replace the Poisson distribution method, which involves the randomness of confirmation delay. After that, \cite{16-Stochastic} summarizes four methods for double-spending. The authors involve the impact of network delay by extending the model proposed in \cite{8-double-spend}. The above double-spending analyses are based on PoW algorithm, and there is not a quantitative analysis to study the double-spending in DAG consensus.

As the most related work, the mathematical analysis of the IOTA White paper \cite{9-tangle} includes three parts: 1) the expected number of tips. 2) cumulative weight growth process of an observed transaction in steady regimes, i.e. HR and LR. 3) probability of the double-spending for large own weight
attack. The author in this work does not consider the impact of unsteady network load in a practical IoT system, which would determine the upper and lower performance bound of the DAG-based ledger. Meanwhile, the analysis for double-spending attack does not consider the impact of adaptation period in DAG consensus process, and does not provide the simplest closed-form expression of the attack success probability.

To this end, we focus on the impact of unsteady network load on DAG consensus process. Compared with the behavior under steady load regimes, we analyse the cumulative weight and confirmation delay under unsteady load regimes to show the upper and lower performance bound of the DAG-based ledger. For the security analysis, we study the most typical double-spending attack in the DAG-based ledger, the parasite chain attack, which refers to the MCMC tips selection algorithm in the practical IOTA system. We consider the adverse impact of adaptation period on security, where the attacker can optimize the strategy by using the computational power of honest nodes. The adverse impact of adaptation period and the attack strategy have not been considered in previous work.

\section{Conclusions and Future Work}

In this work, we use Markov chain model to formulate the consensus process of DAG-based ledger. By identifying four load regimes, our model can capture the dynamic changing of the cumulative weight and the number of tips after a new transaction revealed to the network. Based on the model for DAG consensus process, we leverage a theoretical approach for evaluating the impact of the network load on the
key performance metrics in terms of cumulative weight and confirmation delay with non-attack situation. After that, we involve a typical double-spending attack in consensus process, and use a stochastic model to examine the probability for launching a successful attack under the four load regimes. By conducting numerical simulations, the results demonstrate that the proposed Markov chain model could reflect the features of DAG consensus process under different load regimes accurately, and this can provide an analytical guideline for building optimal and secure DAG-based ledgers in the future.

Compared with PoW and PoS, the impact of network load is a common issue in DAG consensuses (e.g., Tangle,
Byteball and Hashgraph), which has been thoroughly analyzed in this work. However, we cannot directly apply the proposed mathematical models to other DAG consensuses due to the differences in the characteristics among the consensus processes. Nevertheless, the studied problem and designed analysis approach can serve as a foundation for future research of other DAG consensuses. For example, Byteball and Hashgraph have the main chain convergence and famous witnesses election in consensus process respectively. The main chain convergence and famous witnesses election play a key role on system performance and will be directly affected by network load. These topics can be considered as the future work of DAG consensuses.

\begin{IEEEbiography}[{\includegraphics[width=1in,height=1.25in,clip,keepaspectratio]{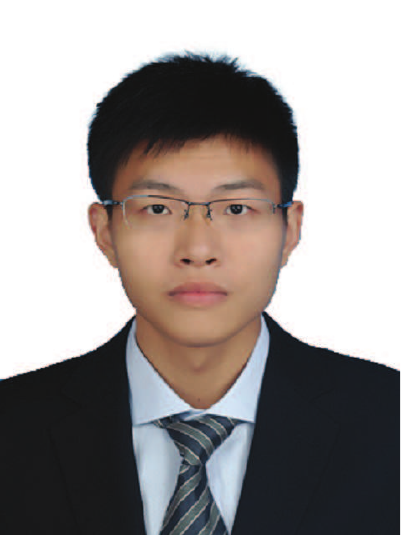}}]{YIXIN LI} 
is pursuing his Master degree at the School of Communication and Information Engineering, Chongqing University of Posts and Telecommunications, Chongqing, China. His research interests include blockchain and internet of things.
\end{IEEEbiography}

\vspace{-6 mm}

\begin{IEEEbiography}[{\includegraphics[width=1in,height=1.25in,clip,keepaspectratio]{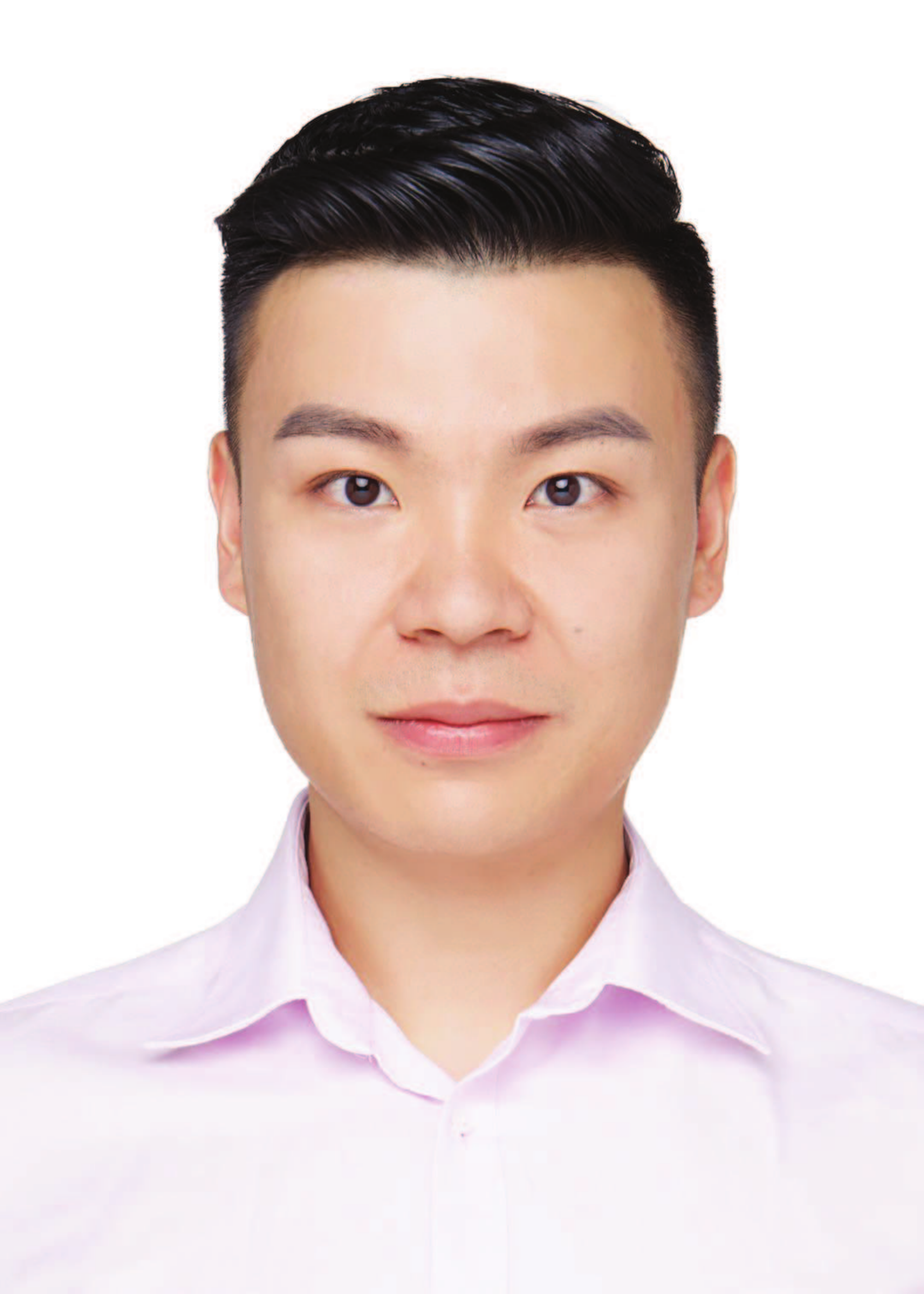}}]{BIN CAO}
is currently an associate professor with the State Key Laboratory of Networking and Switching Technology, Beijing University of Posts and Telecommunications. Before that, he was an associate professor at Chongqing University of Posts and Telecommunications. He received his Ph.D. degree (Honors) in communication and information systems from the National Key Laboratory of Science and Technology on Communications, University of Electronic Science and Technology of China in 2014. From April to December in 2012, he was an international visitor at the Institute for Infocomm Research (I2R), Singapore. He was a research fellow at the National University of Singapore from July 2015 to July 2016. He also served as symposium cochair for IEEE ICNC 2018, workshop cochair for CyberC 2019 and TPC member for numerous conferences. His research interests include blockchain system, internet of things and mobile edge computing.
\end{IEEEbiography}

\vspace{-6 mm}

\begin{IEEEbiography}[{\includegraphics[width=1in,height=1.25in,clip,keepaspectratio]{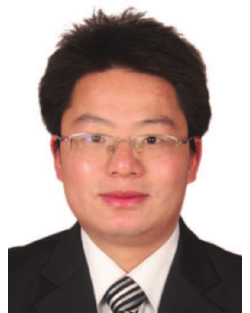}}]{MUGEN PENG} 
received the Ph.D. degree in communication and information systems from the Beijing University of Posts and Telecommunications (BUPT), Beijing, China, in 2005. Afterward, he joined BUPT, where he has been a Full Professor since 2012. His main research areas include wireless communication theory, radio signal processing, cooperative communication, selforganization networking, heterogeneous networking, cloud communication, and Internet of Things. Dr. Peng was a recipient of the 2018 Heinrich Hertz Prize Paper Award, the 2014 IEEE ComSoc AP Outstanding Young Researcher Award, and the Best Paper Award in the JCN 2016, IEEE WCNC 2015, etc.
\end{IEEEbiography}

\vspace{-6 mm}

\begin{IEEEbiography}[{\includegraphics[width=1in,height=1.25in,clip,keepaspectratio]{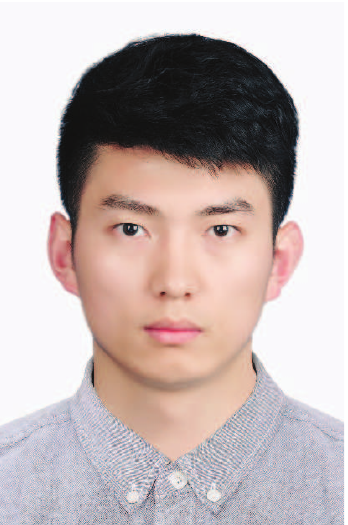}}] {LONG ZHANG} 
received the M.E degree in information and communication engineering from Chongqing University of Posts and Telecommunications, Chongqing, China, in 2019. He currently is pursuing his Ph.D. degree at the National Key Laboratory of Science and Technology on Communications, University of Electronic Science and Technology of China, Chengdu, China. His research interests include mobile edge computing and internet of things.
\end{IEEEbiography}

\vspace{-6 mm}

\begin{IEEEbiography}[{\includegraphics[width=1in,height=1.25in,clip,keepaspectratio]{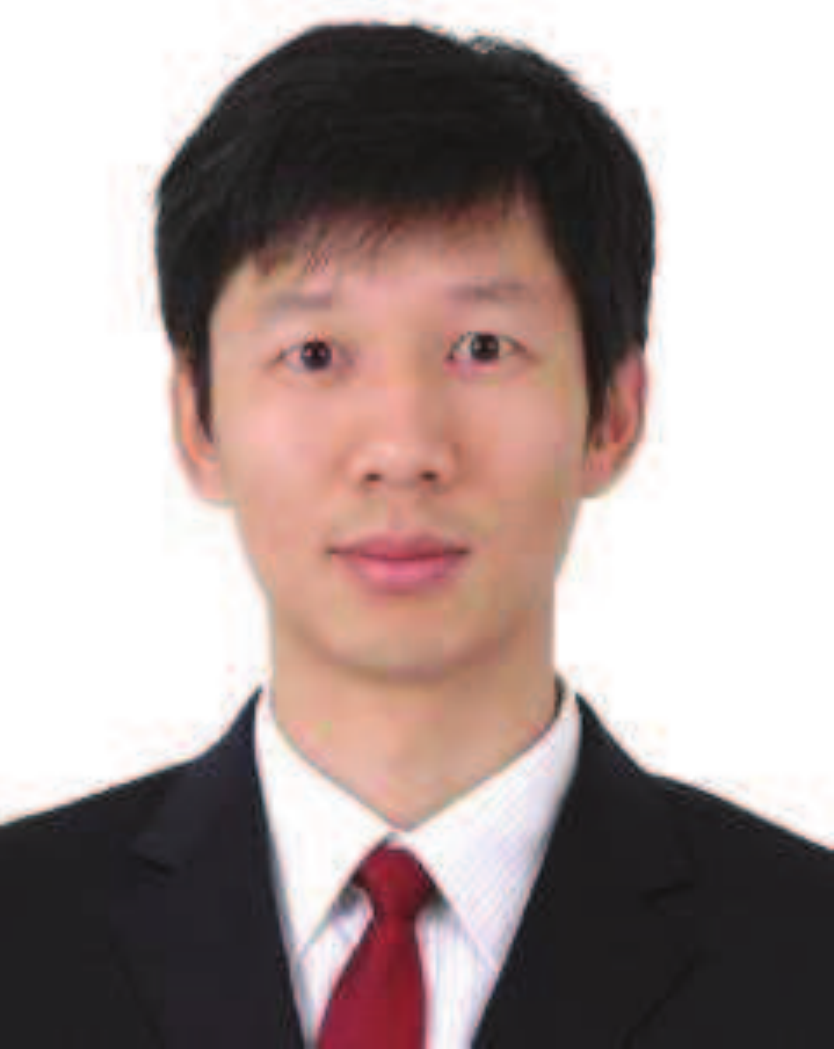}}]{LEI ZHANG} 
received his Ph.D. from the University of Sheffield, U.K. He is now a Lecturer at the University of Glasgow, U.K. His research interests broadly lie in the Communications and Array Signal Processing, including radio access network slicing (RAN slicing), wireless blockchain systems, new air interface design, Internet of Things (IoT), multi-antenna signal processing, massive MIMO systems, etc. He holds 16 US/UK/EU/China granted parents on wireless communications. He also holds a visiting position in 5GIC at the University of Surrey. He is an associate editor of IEEE ACCESS and a senior member of IEEE.
\end{IEEEbiography}

\vspace{-6 mm}

\begin{IEEEbiography}[{\includegraphics[width=1in,height=1.25in,clip,keepaspectratio]{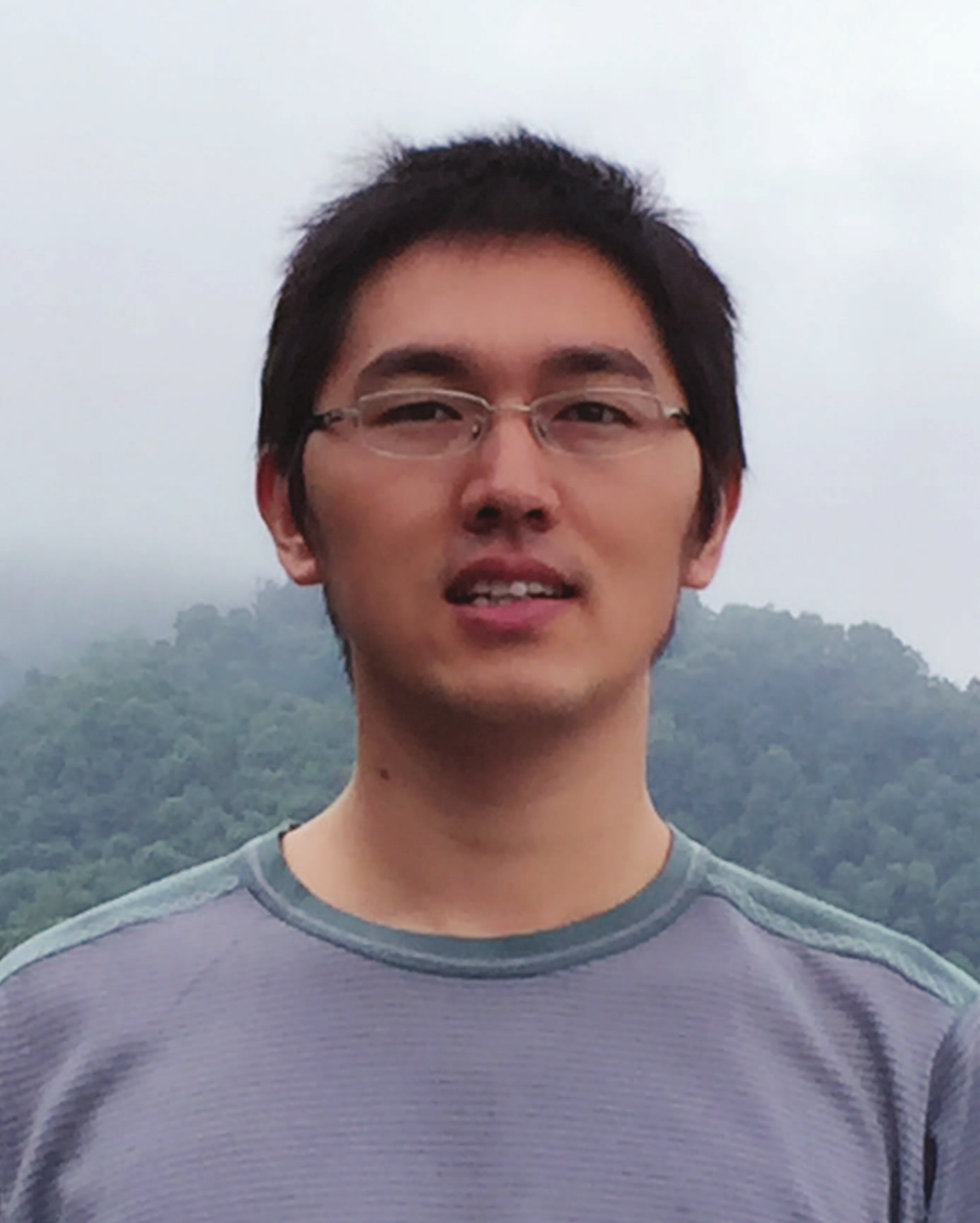}}] {DAQUAN FENG}
received his Ph.D. degree in information engineering from the University of Electronic Science and Technology of China in 2015. He was a Research Staff with State Radio Monitoring Center, Beijing, China, and then a Postdoctoral Research Fellow with the Singapore University of Technology and Design, Singapore. He was a visiting student with the School of Electrical and Computer Engineering, Georgia Institute of Technology, Atlanta, GA, USA, from 2011 to 2014. He is currently an Assistant Professor with the College of Electronics and Information Engineering, Shenzhen University, China. His research interests include URLLC communications, LTE-U, and massive IoT networks.
\end{IEEEbiography}

\vspace{-6 mm}

\begin{IEEEbiography}[{\includegraphics[width=1in,height=1.25in,clip,keepaspectratio]{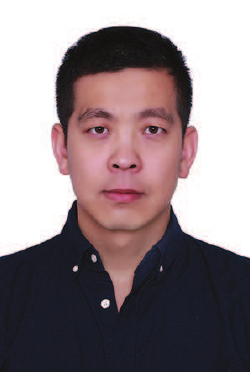}}]{JIHONG YU}
received his Ph.D. degree in computer science at the University of Paris-Sud, Orsay, France, in 2016. He was a postdoc fellow in the School of Computing Science, Simon Fraser University, Canada. He is currently a professor in the School of Information and Electronics at Beijing Institute of Technology.
His research interests include RFID, backscatter networking, and Internet of things.
\end{IEEEbiography}

\end{document}